 \documentclass [12pt,a4paper      ]{article} 
\usepackage{graphics}
\usepackage{times}

\DeclareFontFamily{OT1}{times}{}
\DeclareFontShape {OT1}{times}{m }{n }{ <-> ptmr }{}
\DeclareFontShape {OT1}{times}{bx}{n }{ <-> ptmb }{}
\DeclareFontShape {OT1}{times}{m }{it}{ <-> ptmri}{}
\DeclareFontShape {OT1}{times}{bx}{it}{ <-> ptmbi}{}
\usepackage{amsmath}
\usepackage{amsfonts}
\usepackage{amssymb}
\usepackage{latexsym}
\begin{document}

\title{\bf\vspace{-1.5cm} \emph{Fourth Generation Nuclear Weapons:}\\
                               Military effectiveness and collateral effects}

\author{{\bf Andre Gsponer}\\
\emph{Independent Scientific Research Institute}\\ 
\emph{ Box 30, CH-1211 Geneva-12, Switzerland}}

\date{Version ISRI-05-03.17 ~~ \today}

\maketitle

\begin{abstract}

The paper begins with a general introduction and update to Fourth Generation Nuclear Weapons (FGNW), and then addresses some particularly important military aspects on which there has been only limited public discussion so far.  These aspects concern the unique military characteristics of FGNWs which make them radically different from both nuclear weapons based on previous-generation nuclear-explosives and from conventional weapons based on chemical-explosives: yields in the 1 to 100 tons range, greatly enhanced coupling to targets, possibility to drive powerful shaped-charge jets and forged fragments, enhanced prompt radiation effects, reduced collateral damage and residual radioactivity, etc.

\end{abstract}

\begin{center}
\vspace{3\baselineskip}
\noindent{\Large {\bf Contents}}
\vspace{1\baselineskip}
\end{center}

\renewcommand{\labelitemi}{}   
\renewcommand{\labelitemii}{}
\begin{itemize}

\item \ref{Sec.1}. Introduction

\item \ref{Sec.2}. Second and third generation nuclear weapons
   \begin{itemize}
   \item \ref{Sec.2.1}. Boosting 
   \item \ref{Sec.2.2}. Two-stage thermonuclear weapons 
   \item \ref{Sec.2.3}. Third generation nuclear weapons 
   \item \ref{Sec.2.4}. The ``Robust Nuclear Earth Penetrator'' debate 
   \end{itemize}

\item \ref{Sec.3}. Fourth generation nuclear weapons
   \begin{itemize}
   \item \ref{Sec.3.1}. Inertial confinement fusion experiments and FGNW 
   \item \ref{Sec.3.2}. Microexplosions and high energy-density 
   \item \ref{Sec.3.3}. Petawatt-class lasers (superlasers) 
   \item \ref{Sec.3.4}. Nuclear isomers 
   \item \ref{Sec.3.5}. Antimatter 
   \item \ref{Sec.3.6}. Micro/nano-technology and nuclear bullets 
   \item \ref{Sec.3.7}. Pure antimatter bombs? 
   \end{itemize}

\item \ref{Sec.4}. Target coupling
   \begin{itemize}
   \item \ref{Sec.4.1}. Initial energy from conventional or nuclear weapons 
   \item \ref{Sec.4.2}. Initial work from conventional or nuclear weapons 
   \item \ref{Sec.4.3}. Coupling to homogeneous and heterogeneous targets 
   \item \ref{Sec.4.4}. FGNW coupling 
   \end{itemize}

\item \ref{Sec.5}. Thermonuclear-driven jets and projectiles
   \begin{itemize}
   \item \ref{Sec.5.1}. Conventional shaped-charges 
   \item \ref{Sec.5.2}. ``Nuclear'' and ``thermonuclear'' shaped-charges 
   \item \ref{Sec.5.3}. FGNW-driven jets and projectiles 
   \end{itemize}

\item \ref{Sec.6}. Collateral effects
   \begin{itemize}
   \item \ref{Sec.6.1}. Mechanical and thermal effects 
   \item \ref{Sec.6.2}. Prompt radiation effects 
   \item \ref{Sec.6.3}. Delayed radiological effects
   \item \ref{Sec.6.4}. Electromagnetic effects  
   \end{itemize}

\item \ref{Sec.7}. Conclusion
   \begin{itemize}
   \item \ref{Sec.7.1}. Military aspects 
   \item \ref{Sec.7.2}. Technical aspects 
   \item \ref{Sec.7.3}. Political aspectss 
   \end{itemize}

\item \rule{0mm}{1mm}~~~~~Acknowledgements 

\item \rule{0mm}{1mm}~~~~~References 
 
\end{itemize}
\renewcommand{\labelitemi}{$\bullet$}    
\renewcommand{\labelitemii}{\bfseries --}

\section{Introduction}
\label{Sec.1}

   Sixty years after the only use of nuclear weapons in warfare, this paper is discussing the elaboration and characteristics of a forthcoming generation of war-fighting nuclear weapons which has been under serious consideration for more than fifty years, and which may become a reality within a decade or two.

Many technical and some political aspects of these \emph {fourth-generation nuclear weapons} (FGNW) have been discussed in earlier reports \cite{GSPON1997A, GSPON1998-}.  The present paper will recall some of them, while focusing on the military effectiveness and collateral effects of the concepts which can currently be perceived as the most likely to move from the laboratory to the battle-field.

The paper will therefore not address the full spectrum of possibilities that exist for designing new types of nuclear weapons (which in the professional scientific literature comprises many concepts, including very hypothetical ones such as ``quark'' or ``black hole'' bombs) but only truly realistic design which are at most a few decades away from being usable on the battle-field.  Neither will the paper discuss more conservative and near-term possibilities based on the refinement of existings weapons, such as ``simple, rugged designs'' \cite[p.11]{YOUNG2000-}, or ``robust earth penetrators'' \cite{GSPON2003A}, even though some comments will be made when useful to put them in perspective with the development of FGNWs.

In particular, we will specialize to explosive devices in which the main source of militarily useful energy (i.e., yield) is the deuterium-tritium ($DT$) fusion reaction.  As will be seen, when used as the dominant process in a nuclear weapon, this reaction leads to very different military effects than those of first and second generation nuclear weapons (i.e., ``A'' and ``H'' bombs), where \emph{in fact} the dominant source of yield is fission, and where fusion plays only a secondary role in enhancing the fissioning of large amounts a fissionable materials.

This means that the discussion of FGNWs requires some intellectual effort --- especially for non-technical minded people --- because FGNWs are in many ways very different from previous generation weapons.  This can be illustrated by comparing a few salient features of typical first and fourth generation explosives:

  \begin{center}
  \fbox{First ~ generation: ~ {\bf ~~ ~~6 kg $Pu$} ~~ $ ~~ \approx ~~ $ ~~ 10~\emph{kt} ~ yield at 10\% efficiency}
  \end{center}
  \begin{center}
  \fbox{Fourth generation: ~ {\bf 25 mg $DT$} ~~ $ ~~ \approx ~~ $ ~~ 1~\emph{ton} ~ yield at 50\% efficiency}
  \end{center}

Consequently, going from the first to the fourth generation implies a total change of perspective about nuclear weapons:  A ``change of paradigm'' where the concept of very-large-yield and big nuclear weapons for deterrence-use is shifting towards the concept of very-high-precision and compact nuclear weapons for battle-field-use --- with yields in the 1 to 100 \emph{tons}\footnote{We use italics when \emph{tons}, \emph{kt}, or \emph{Mt} refer to tons of TNT equivalent.} range, that is intermediate between conventional and contemporary nuclear weapons.

Moreover, another important paradigm shift \cite[Endnote~39]{GSPON1998-} is the move towards ``virtual nuclear weapons'' and ``virtual nuclear weapon States'' \cite{MAZAR1995-}, as well as to ``factory deterrence'' \cite{WALTZ1997-}, ``technical deterrence,'' or ``deterrence by competence'' \cite{YOUNG1997-}.  This is because FGNWs do not need to be actually built and deployed before that can play significant strategic and political roles.  What matters, as in the case of the proliferation of so-called ``peaceful nuclear energy,'' is technological deterrence, that is the ability to master the technology and to apply it if necessary to military ends.

Finally, as can be seen by the references given in the bibliography, this paper is an opportunity to make a synthesis of much of the work done at ISRI since 1982 (the year of its creation) on third and fourth generation nuclear weapons.  This does not mean that there are no other references to consult: quite the contrary, as can be seen from the numerous references cited in these ISRI documents.  Similarly, there will be a special emphasis on the use of microscopic amounts of antimatter to trigger FGNWs, even though other potent ignition technologies are concurrently under development. The main reason for this is that antimatter is the most promising technology for achieving the extremely high energy-densities required to ignite thermonuclear detonation-waves in very compact thermonuclear explosives.

\section{Second and third generation nuclear weapons}
\label{Sec.2}

   Almost all nuclear weapons which are currently in the arsenals of official and \emph{de facto} nuclear-weapons-states are so-called ``second generation nuclear weapons.''  In these weapons a small amount of tritium (a few grams, in the form of $DT$ gas) is used to insure the reliability and safety of the nuclear fission-explosives, which can be used on their own (``boosted fission bombs''), or as primaries of two-stage thermonuclear weapons (``hydrogen bombs'').  Weapons which in contemporary arsenals do not use ``tritium boosting'' have generally sub- or low-\emph{kiloton} yields, and are mostly special weapons such as atomic demolition munitions.

\subsection{Boosting}
\label{Sec.2.1}

   Tritium boosting, which in practice consists of imploding thin hollow fissile-material spheres containing $DT$ gas, is relatively easy to implement. This is because the high-explosive technique required for that purpose is the same as the one used for imploding metallic liners in anti-tank shaped-charges \cite{DELON1995-}.  Tritium boosting has a number of significant technical and military advantages, which explain why it is used in essentially all militarized nuclear weapons, including in India, Pakistan, and North-Korea.  These advantages comprise:
\begin{itemize}

\item high-efficiency with relatively low-compression and thin reflector/tamper;

\item low weight and small size;

\item intrinsic-safety capability (zero or negligible yield when the tritium is not in the weapon\footnote{The boost gas has a small neutron moderating effect which can be exploited to design a weapon in such a way that a diverging chain reaction stops at an early stage when no boost gas is present in the pit.  This is difficult to achieve in practice and required many tests in the past, in particular to ensure ``one-point safety'' (which in fact can only be secured when there is no boost gas in the pit).});

\item preinitiation-proof capability (resistance to neutrons from spontaneous-fission or other warheads);

\item high transparence to X-rays.

\end{itemize}

    The last advantage derives from the fact that a boosted device does not need a thick and heavy neutron-reflector/tamper to insure a sufficient yield, so that most of the energy released by the neutron chain-reaction (which at the end of the fission explosion is found in the form of hard-X-rays) can easily escape from the fissioned material.  The first effect of these X-rays is to heat the surrounding materials, such as the compressed reflector/tamper and the  detonation products of the high-explosive that imploded these materials, as well as the shell of heavy material (e.g., steel or uranium  -- sometimes called the ``barrel shell'') which is generally used to contain the high-explosives.  Consequently, if this outer-shell is sufficiently thick and opaque to X-rays (e.g., 1 cm of uranium), it will become extremely hot and therefore behave as a secondary-source of X-rays: This is the principle of the ``hot soft-X-ray source'' which is used as the ``primary'' of a two-stage H-bomb.\footnote{Since the diameter of this outer-shell is about ten times the diameter of the compressed fission material, the duration of the soft-X-ray pulse from the barrel shell will also be about ten times the duration of the X-ray pulse from the fissioned material --- an important consideration if the pulse is to compress the fusion material of an H-bomb secondary.}

\subsection{Two-stage thermonuclear weapons}
\label{Sec.2.2}

\begin{figure}
\begin{center}
\resizebox{15cm}{!}{ \includegraphics{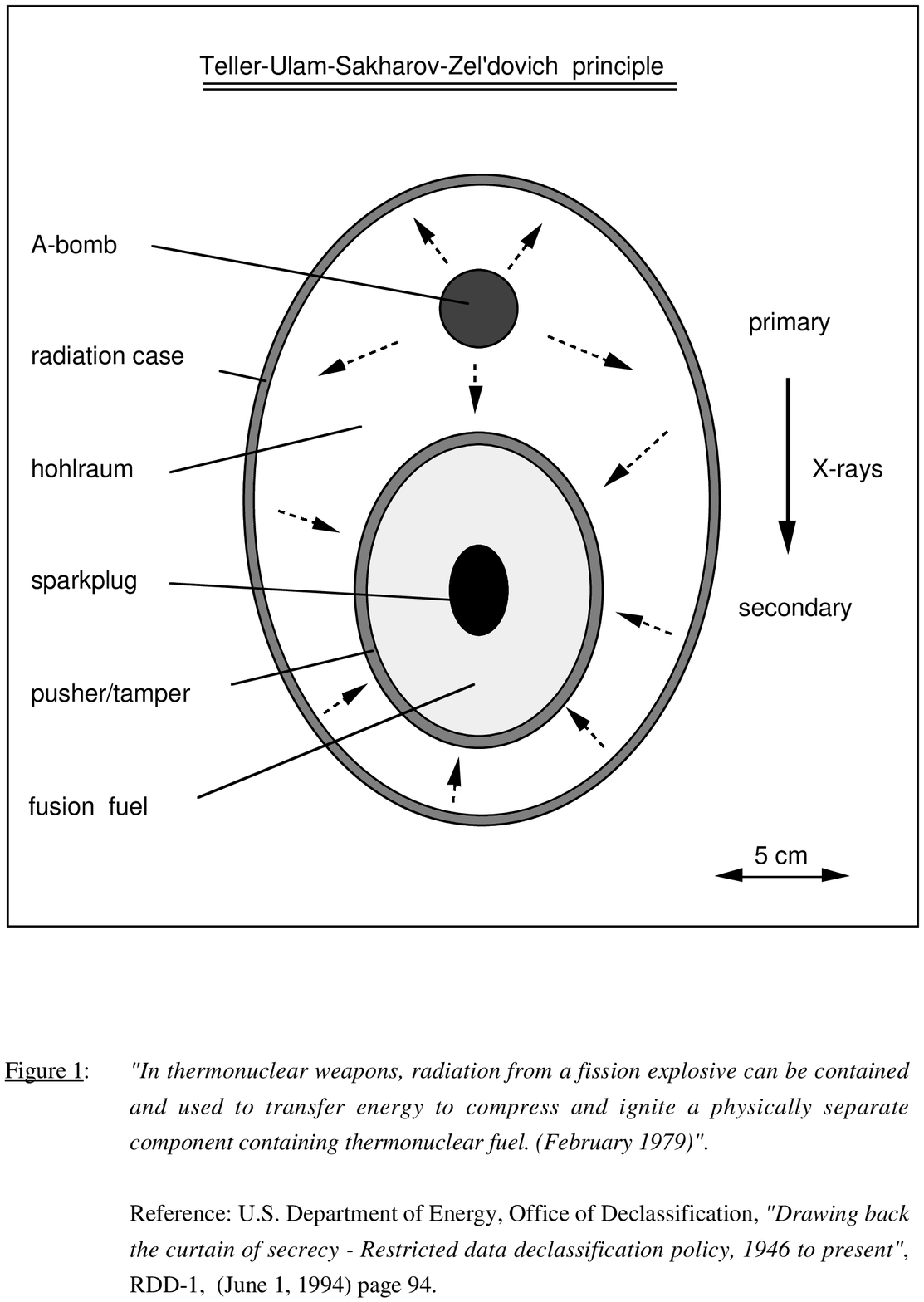}}
\end{center}
\caption{Fig.~3 of Ref.~\cite{GSPON1997A}}
\label{fig:TUS}
\end{figure}

   In two-stage thermonuclear weapons, the fusion material (i.e., lithium-deuteride, $LiD$) is generally packaged as a cylindrical or spherical shell sandwiched between an outer-shell of heavy material (the pusher/tamper) and an inner-shell of fissile material.  This inner-shell (the spark-plug) is generally boosted with some $DT$ gas.  As suggested by its name, the purpose of the spark-plug is to ignite the fusion material at the appropriate time, i.e, once it has been sufficiently compressed.  This whole package is call the ``secondary'' of the thermonuclear weapon, and is enclosed together with the ``primary'' in a thick and heavy ``radiation case'' which is designed to contain the soft-X-rays from the primary as long as possible (See Fig.~\ref{fig:TUS}).

   The explosion proceeds as follows:  The soft-X-rays from the primary fill the radiation case and start interacting with the outer casing of the secondary (the pusher/tamper).  The material at the surface of the pusher/tamper is ablated under the effect of these X-rays, and by reaction creates an enormous pressure that starts compressing the whole secondary.  When the compression is such that fission reactions can start in the spark-plug, sufficient heat can be generated at the center of the secondary so that fusion reactions can start as well in the compressed fusion material.  Under suitable conditions the fusion reaction can continue on its own, until a substantial fraction of the fusion fuel is burnt, generating large amounts of energy and high-energy fusion-neutrons.

  To further increase the energy of the H-bomb, the fusion neutrons are generally used to induce fission reactions in the pusher/tamper, and/or in the radiation case, which for that purpose have to be made out of natural or enriched uranium, or out of low-grade plutonium.\footnote{Any plutonium (even reactor grade) can be used for that purpose, especially if it is just used for the part of the radiation case that surrounds the secondary.  This is a major nuclear weapons proliferation problem associated with separated plutonium.  The use of reactor grade plutonium in its thermonuclear weapons is apparently the option chosen by India \cite{NATUR1998-}.} As a result, one get a ``third stage'' which in most thermonuclear weapons provides the main contribution to the total yield.  Consequently, in contemporary thermonuclear weapons, most of the yield actually comes from fission rather than fusion, as can be seen in Table~\ref{tbl:1}.

\begin{table}
\begin{center}
\begin{tabular}{|c|c|c|c|}

\hline
\multicolumn{4}{|c|}{\raisebox{+0.4em}{{\bf Fission and fusion fractions\rule{0mm}{6mm}}}} \\
 \hline
          
{\bf ~}             &   Yield  &   Fission  &    Fusion    \rule{0mm}{5mm}\\

~                   &   [\emph{kt}]   &    [\%]    &     [\%]     \\

\hline

Mike               &  10400    &     77     &     23      \rule{0mm}{5mm}\\

\hline

B-28               &   800     &    80      &     20      \rule{0mm}{5mm}\\

\hline



W-88               &   325     &     70     &     30      \rule{0mm}{5mm}\\

\hline

W-88               &   475     &     80     &     20      \rule{0mm}{5mm}\\

\hline
\end{tabular}

\caption{\emph{Estimated fission and fusion fractions in the first (Mike) and latest (W-88) United States's thermonuclear weapons.}}
\label{tbl:1}
\end{center}
\end{table}

  To conclude this brief introduction to second generation nuclear weapons, let us summarize their main characteristics:

\begin{itemize}

\item Most of the yield of thermonuclear weapons is produced in a ``third stage'' which is fissioned by the fusion neutrons leaking out from the second stage.  At the end of the thermonuclear explosion the weapon is therefore primarily a very powerful source of X-rays.

\item The ``50\% fission, 50\% fusion'' estimate used in unclassified fallout calculations is optimistic: thermonuclear weapons are actually more dirty.

\item First and second generation nuclear weapons are primarily gigantic sources of soft-X-rays generating heat and blast.  The fission neutrons leaking out from the weapons have a comparatively smaller effect.

\item The yields are typically very large, on the order of \emph{kilotons} to \emph{megatons}, essentially because of the ``tyranny of the critical mass'' which implies that fission-based explosions will normally have yields in the \emph{kiloton} range or higher.

\end{itemize}

Finally, as a transition to the next subsection, it is important to remark that a two-stage H-bomb demonstrates that a powerful source of X-rays can be used to produce mechanical work, i.e., to strongly compress the material of the secondary.  This leads to other possible applications, where the ablation pressure is used to accelerate a missile or a spacecraft (nuclear-driven rocket), or to squeeze a shaped-charge liner (nuclear-driven plasma-jet).

\subsection{Third generation nuclear weapons}
\label{Sec.2.3}

  Third generation nuclear weapons are basically ``tailored and special effects'' warheads and systems developed between the 1960s and 1980s, mainly for tactical uses or ballistic missile defense. Examples of these developments comprise the following concepts:

\begin{itemize}

\item ERW --- Enhanced Radiation (neutrons, hard X-rays) 

\item RRR --- Reduced Residual Radiation (enhanced blast)

\item EMP --- enhanced ElectroMagnetic Pulse

\item DEW --- Directed Energy (plasma-jet or X-ray laser-beam)

\item EPW --- Earth Penetrating Warhead

\item ETC --- {\bf ...}

\end{itemize}

Most of these weapons were never deployed on a large scale for a number of technical and political reasons. In particular, third generation nuclear weapons require a fission-explosive as trigger, which implies that their yield tends to be too high for battle-field uses, and that they necessarily produce large-scale radioactive pollution, etc.

As will be seen in the sequel of this paper, most third generation concepts can be reconsidered in the context of fourth generation nuclear weapons. This is because the suppression of the fission-explosive trigger, and the reliance on fusion rather than fission as the main source of yield in FGNWs, enable to envisage devices of much lower yield and much reduced radiological impact.

\subsection{The ``Robust Nuclear Earth Penetrator'' debate}
\label{Sec.2.4}

The limitations in the military usefulness of third generation concepts, as well as the relations of third- to fourth-generation nuclear weapons, are well illustrated by the on-going debate on the development by the United States of an improved EPW version of the B61, the ``Robust Nuclear Earth Penetrator'' \cite{GSPON2003A}.

  The current version, B61-11 (that is the 11th modified version of the 100 to 500 \emph{kt} B61 gravity bomb whose development started in 1961), entered service in 1997.  It has a penetration capability of a few meters in solid rock or reinforced concrete.  In the new version, this range would be extended to about 15 to 30 meters.  This leads to three remarks:

\begin{enumerate}

\item  If the maximum penetration range is extended from 6 to 30 meters, which is a formidable technical challenge, the military effectiveness of the weapon will only be \emph{marginally} improved.  This is because the B61 cannot destroy targets at depths or distances much larger than 30 to 100 meters below or away from the point of explosion.  Therefore, the only way to have a truly more effective weapon would be to design a third-generation tandem-warhead device, such as a ``DEW plus EPW'' combination, which appears to bo both very complicated and difficult to develop without extensive nuclear testing.  On the other hand, a FGNW tandem-warhead system with a large penetration capability could possibly be more feasible and easier to develop.

\item  Any nuclear weapon detonated below the surface of the Earth creates a radioactive hot spot, containing the highly-radioactive fission-products, as well as the fissile materials that have not been fissioned during the explosion --- that is more than 90\% of the initial fissile-materials content of the warhead.\footnote{In the case of an air-burst, the fission products and non-fissioned fissile materials are dispersed over a large surface or carried away in the form of nuclear fallout.}  This leads to numerous problems, especially if the weapon contained plutonium.  Indeed, most of the un-fissioned plutonium left in the cavity created by an underground nuclear explosion can later be recovered, possibly by terrorists, and reused to manufacture nuclear weapons.  This is the problem of \emph{``plutonium mines,''} which is a major nuclear proliferation concern in countries such as Kazakhstan \cite{STONE2003-} and Algeria, where numerous plutonium-based nuclear explosives were detonated underground.  

Consequently, earth penetrating warheads like the B61 use highly-enriched uranium as fissile material, because in that case the un-fissioned U-235 cannot be recovered after the explosion as it will be intimately mixed with the U-238 from the non-enriched uranium components of the weapon.

  On the other hand, the un-burnt fraction of the tritium used for boosting is less of a problem.  First because tritium is very volatile and easily drifts away from the point of explosion, second because its half-life is only twelve years, and third because its radio-toxicity is relatively small.  For these reasons, in the case of FGNWs, the related problem of ``tritium mines'' is potentially less serious than the problem of ``plutonium mines.''

\item  To extend the penetration range of the B61-11, the crucial improvements to be made are not in the major nuclear components of either the primary or the secondary (which are intrinsically rugged, see Ref.~\cite{GSPON2003A}) but rather in the penetrator casing, and in the nuclear package's ancillary components, which in any warhead are the most delicate and fragile ones: The fuzeing and initiation systems, the high-explosive detonators, the neutron generators, the safety and security devices, etc. It is primarily the resistance to failure of these key components that has to be improved by order of magnitude factors to make them compatible with the dramatically increased stresses stemming from the higher impact velocities required to achieve deeper penetration.  

Some of these improvements are directly linked to advances in the fields of \emph{micro-electro\-mechanical engineering and nanotechnology} \cite{GSPON2002B}.  Since these potential advances significantly overlap with the requirements of many future weapons, both conventional and nuclear, it is possible that an important reason for considering an improved version of the B61-11 is simply to develop these technologies as they will be needed, for instance, to weaponize FGNWs.

\end{enumerate}

\section{Fourth generation nuclear weapons}
\label{Sec.3}

   There is no standard definition of fourth generation nuclear weapons.  Nevertheless, for the purpose of this paper, we may use either of the two definitions:

\begin{itemize}

  \item \emph{``Nuclear explosive devices based on atomic and nuclear processes that are not restricted\,\footnote{The fact that the development and deployment of fusion-based FGNWs is not restricted by the CTBT, and that such weapons may also be developed and deployed by non-nuclear weapon States, is discussed in Chap.~2 of Ref.~\cite{GSPON1997A}.} by the Comprehensive Test Ban Treaty (CTBT),''} or

  \item \emph{``Nuclear explosive devices based on low-yield thermonuclear pellets triggered by compact non-fission primaries.''}

\end{itemize}

The second definition recognizes the technical fact that radically new, but \emph{realistic}, types of nuclear weapons will most probably use highly-compressed deuterium-tritium pellets as the main source of their explosive energy.  This means that while fission was the main source of yield in the first three generations, the main source of yield in the fourth generation will be the fusion reaction
\begin{equation}\label{DT-reaction}
D + T \rightarrow  He^{4} ~(3 \text{~MeV})  +  n ~(14 \text{~MeV}) .
\end{equation}

The physics of the ignition and burn of such $DT$ pellets is vigorously studies in all nuclear weapons states, as well as in a few other technologically advanced countries, most prominently in Japan and in Germany \cite{GSPON1997A, GSPON1998-}.

\subsection{Inertial confinement fusion experiments and FGNW}
\label{Sec.3.1}

\begin{figure}
\begin{center}
\resizebox{15cm}{!}{ \includegraphics{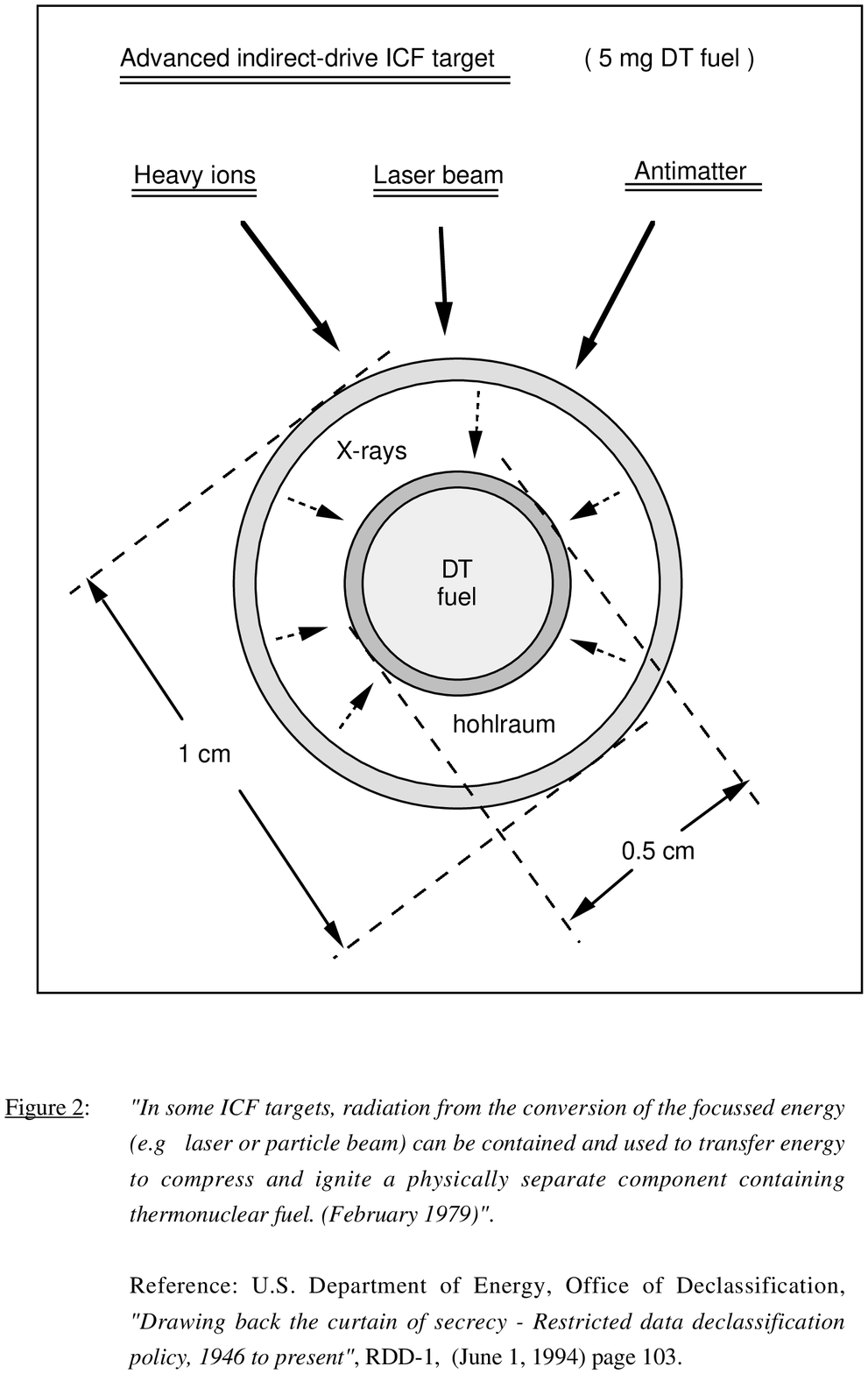}}
\end{center}
\caption{Fig.~13 of Ref.~\cite{GSPON1997A}}
\label{fig:ICF}
\end{figure}

Inertial confinement fusion (ICF), which basically consists of exploding very small amounts of thermonuclear fuel highly compressed by lasers or other means, enables to study the physics of thermonuclear secondaries in the laboratory (see Fig.~\ref{fig:ICF}).  While this technique has the potential to be used in a thermonuclear-reactor to produce energy, it has primarily been developed as an alternative to the underground testing of nuclear weapons, and as a tool for designing new types of nuclear weapons \cite{GSPON1997A}.

The world's two largest ICF facilities are presently under construction in the USA at Livermore (NIF)\footnote{For a recent appreciation of the progress on NIF see Ref.~\cite{MOSES2005A}.} and in France at Bordeau (LMJ). However, the most powerful ICF facility presently in operation is in Japan at Osaka (ILE), with ambitious plans to upgrade it.\footnote{This was written before the author became aware of the latest progress made with the construction of NIF, where in August 2005 the combined output of eight (out of 192) beams totaled 152 kilojoules, exceeding design specifications and becoming the higest ever reached energy output of a pulsed laser \cite{MOSES2005B}.}  Another powerful and flexible facility presently under construction is in Germany at Darmstadt (GSI), which has the effect to put Germany in the same ``club'' as the other most advanced countries in this field, i.e., the United States, France, Japan, and the United Kingdom. 

Basically, as can be seen by comparing Fig.~\ref{fig:TUS} and \ref{fig:ICF}, ICF reproduces in the laboratory the same arrangement than the one on which two-stage H-bomb are based: the ``Teller-Ulam principle.''  Provided the compression and ignition of thermonuclear pellets is well understood, and yield in excess of a few \emph{tons} can be obtained by simple and reliable means, there is a possibility of using these pellets as the main explosive of a FGNW.

In practice, according to reasonable extrapolations of the present understanding of the physics of ICF, it is very likely that ignition and burn of ICF pellets will be fully understood within the next decade or so, with the help of experiments made at NIF, LMJ, and other facilities.\footnote{For a recent review of ICF physics see Ref.~\cite{LINDL2004-}.}  Moreover, advanced pellets containing about 5~mg of $DT$ such as depicted in Fig.~\ref{fig:ICF} and~\ref{fig:ADV}, which are considered for use in future ICF power plants, will have yields on the order of 0.1~\emph{ton}, assuming a burn efficiency of 25\%.

In order to use these pellets in a FGNW, the remaining problem --- which of course is a formidable technical challenge --- will be to design a compact trigger, i.e., a non-fission primary, to compress and ignite them in a weaponizable configuration.  For this purpose, as is indicated at the top and bottom of Fig.~\ref{fig:ADV}, some of the most attractive technologies are superlasers, nuclear isomers, and antimatter.  Before discussing these technologies, let us recall why such ``high-energy-density'' materials and processes are required.

\subsection{Microexplosions and high energy-density}
\label{Sec.3.2}

Let us consider a sphere of fission or fusion material of radius $R$ and density  $\rho > \rho_0$, its normal density.  If this sphere is totally fissioned or fused, one will get a maximum possible yield $Y_{max}$ proportional to its volume and density.  Therefore, if one introduces a burn efficiency $\eta$, the yield can be written as: 
\begin{equation}
  Y = \eta Y_{max}  \text{~~~ where ~~~} Y_{max} \propto  \rho R^3  .
\end{equation}

   In the case of a fissionable material, a first approximation formula for the efficiency is given by the expression
\begin{equation}\label{fis-eff}
  \eta_{fission} \approx \frac{\rho R - \omega_c}{\rho R} ,
\end{equation}
where the parameter $\omega_c$ is called the ``critical fast-neutron-opacity'' ($\omega_c \approx 100$~g/cm$^2$ for $^{239}Pu$, and 160~g/cm$^2$ for $^{235}U$).  As can be seen, there is a minus sign in this expression which means that there will be some yield only if the product $\rho R$, called ``opacity,'' is larger than $\omega_c$:  this is the mathematical translation of the fact that with fission there is a critical mass below which (at any given density) a fission chain reaction is not possible.

   In the case of a fusionable material, the efficiency is given in good approximation by the expression
\begin{equation}\label{fus-eff}
   \eta_{fusion} \approx \frac{\rho R}{\rho R + B} ,
\end{equation}
where the parameter $B$ is called the ``burn parameter'' ($B \approx 6$~g/cm$^2$ for $DT$, and 17~g/cm$^2$ for $^{6}Li_2DT$).  In that case, there is no critical mass and there is some yield at any given density.

   There are therefore two intrinsic advantages in fusion over fission explosives: there is no critical mass, and there can be some yield at any given density, i.e., at any compression of the material.\footnote{This implies that the yield of FGNWs will be easily and continuously adjustable, contrary to A and H-bombs in which yield-adjustment is difficult.}  However, in either fission or fusion, it is desirable that the efficiency is as close as possible to $\eta = 1$.  Looking at equations \eqref{fis-eff} and \eqref{fus-eff}, this implies that the product  $\rho R$ has to be as large as possible, meaning that for a given amount of material the density has to be as large as possible.  Thus, if high-efficiency is desired at small yields (i.e., small $R$), it is necessary that the density $\rho$ be very high.

   But equations \eqref{fis-eff} and \eqref{fus-eff} are quasi-static approximations that are only valid if the fission or fusion material can be considered as being instantly compressed to the density $\rho$, and if the fission or fusion reaction starts at that moment only.  Indeed, many practical problems arise during compression, such as instabilities and preignition, which can only be alleviated if the implosion is very fast, i.e., the implosion velocity very high.

   To find the key parameter leading to both high density and high implosion-velocity, let us recall the main features of any explosive process, e.g., chemical reaction, nuclear reaction, X-ray-driven ablation, etc.  They are
\begin{align}
\text{Maximum pressure:}    ~~ ~~ ~~   p  &\propto   Q,\\
\text{Detonation velocity:}  ~~ ~~ ~~   D  &\propto  \sqrt{Q/\rho},
\end{align}
\noindent where $Q$, [J/m$^3$], is the energy density characteristic of the process.  Then, for a material under very high pressure, assuming the Thomas-Fermi approximation, one has
\begin{align}
\text{Density:}  ~~ ~~ ~~  \rho \propto p^{3/5} ~~ \Rightarrow ~~ \rho  &\propto  Q^{3/5},\\
\text{Implosion velocity:}  ~~ ~~ ~~   v \approx D~~ \Rightarrow ~~  v  &\propto  Q^{1/5}.
\end{align}
Therefore, one can conclude that the key parameter is the maximum energy-density provided by the material or process used to compress the fission or fusion fuel to be ignited and burnt.  It follows that the development of nuclear and thermonuclear explosives is directly correlated to that of high energy-density techniques and materials, of which the most important ones are compared in Table~\ref{tbl:2}.

\begin{table}
\begin{center}
\begin{tabular}{|l|r|r|}

\hline
\multicolumn{3}{|c|}{\raisebox{+0.4em}{{\bf High energy-density techniques and materials\rule{0mm}{6mm}}}} \\
\hline
Electric energy  &  0.01 &  \rule{0mm}{5mm} kJ/cm$^3$ ~~~~~~\\

Magnetic energy  & 0.1 & kJ/cm$^3$ ~~~~~~\\

Kinetic energy   & 1 & kJ/cm$^3$ ~~~~~~\\

Chemical energy  & 10 & kJ/cm$^3$ ~~~~~~\\

\hline

Magnetic compression &  0.1--1 & \rule{0mm}{5mm} MJ/cm$^3$ ~~~~~~\\

Laser (NIF, LMJ)     &   1--10 &  MJ/cm$^3$ ~~~~~~\\
  
Soft X-rays***       &     420 &  MJ/cm$^3$ ~~~~~~\\

Nuclear isomers**    & 1--1000 &  MJ/cm$^3$ ~~~~~~\\

\hline

Fusion energy     &  100   & \rule{0mm}{5mm}  GJ/cm$^3$ ~~~~~~\\

Fission energy    & 1000   & GJ/cm$^3$ ~~~~~~\\

Annihilation energy*  &  10000 & GJ/cm$^3$ ~~~~~~\\

\hline
\multicolumn{3}{c}{\raisebox{+0.1em}{ *** 1~\emph{kt} in 10~dm$^3$   \rule{0mm}{6mm}}}\\
\multicolumn{3}{c}{\raisebox{+0.1em}{ ** Not yet demonstrated  \rule{0mm}{2mm}}}\\
\multicolumn{3}{c}{\raisebox{+0.1em}{ * No fundamental problem \rule{0mm}{2mm}}}\\
\end{tabular}
\caption{\emph{Typical maximum energy densities provided by various physical processes or materials.}}
\label{tbl:2}
\end{center}
\end{table}

In order to interpret this table let us first consider what could be achieved if a chemical high-explosive such as RDX was used to compress $LiD$, the fusion fuel used in the secondary of H-bombs.  The approximate characteristics of RDX are: $Q =  10$~kJ/cm$^3$, $D =  0.8$~cm/$\mu$s, and $p = 0.4$~Mbar $= 40$~GPa.  Assuming that one does nothing complicated (i.e., no shock convergence, no ``levitation,'' etc.), the maximum static compression can be found by looking up standard equation of state tables, which give $ \rho / \rho_0 \approx  2.2$.  Assuming $\rho_0 = 0.9$~g/cm$^3$, equation \eqref{fus-eff} gives an efficiency $\eta_{fusion} = 0.5$ for $R \approx 8.5$~cm.  This corresponds to about 5~kg of $LiD$, and therefore to 125~\emph{kt} of yield if that fuel could be ignited under these conditions.  Consequently, if a smaller yield is desired, a much higher energy-density than provided by RDX is required to achieve the correspondingly larger compression.

The same conclusion is reached if one realizes that the 5~kg of $LiD$ of that example is on the order of the typical amount used in a contemporary thermonuclear weapon secondary.  In such a weapon the energy-density of the X-rays required to implode the secondary is $\approx 1$~\emph{kt} in a volume of $\approx 10$~dm$^3$, i.e., 420~MJ/cm$^3$, that is 42,000 times more than the energy density provided by RDX. 

It is thus out of question to directly use chemical explosives to compress a small pellet of thermonuclear fuel and expect a reasonable efficiency:  this requires at least a powerful laser such as NIF or LMJ, or else a novel technology such as magnetic compression \cite{YOUNG1996-,CHERN1996-,DEGNA1999-}, nuclear isomers, or antimatter.

An interesting aspect of this table is that all techniques which have energy densities in the MJ/cm$^3$ range provide this energy in the form of electromagnetic waves, i.e., ``photon gases'' of increasing temperatures: about 1~eV for lasers, 10--100~eV for magnetized fluids, 1~keV for soft X-rays, 1~MeV for nuclear isomers.  Consequently, the basic concept is always somehow related to the Teller-Ulam principle: compressing a secondary by means of a hot photon gas generated by a primary.

In conclusion, we confirm what is indicated in Fig.~\ref{fig:ADV}, namely that some of the most promising techniques for compressing and igniting ICF pellets are high-power lasers, nuclear isomers, and antimatter.

\subsection{Petawatt-class lasers (superlasers)}
\label{Sec.3.3} 

As was explained in the previous sections, the two main issues in exploding a thermonuclear pellet are (i) to implode the pellet to high density as fast as possible, and (ii) to ignite the pellet by heating it at the moment it has reached maximum compression.

\begin{figure}
\begin{center}
\resizebox{15cm}{!}{ \includegraphics{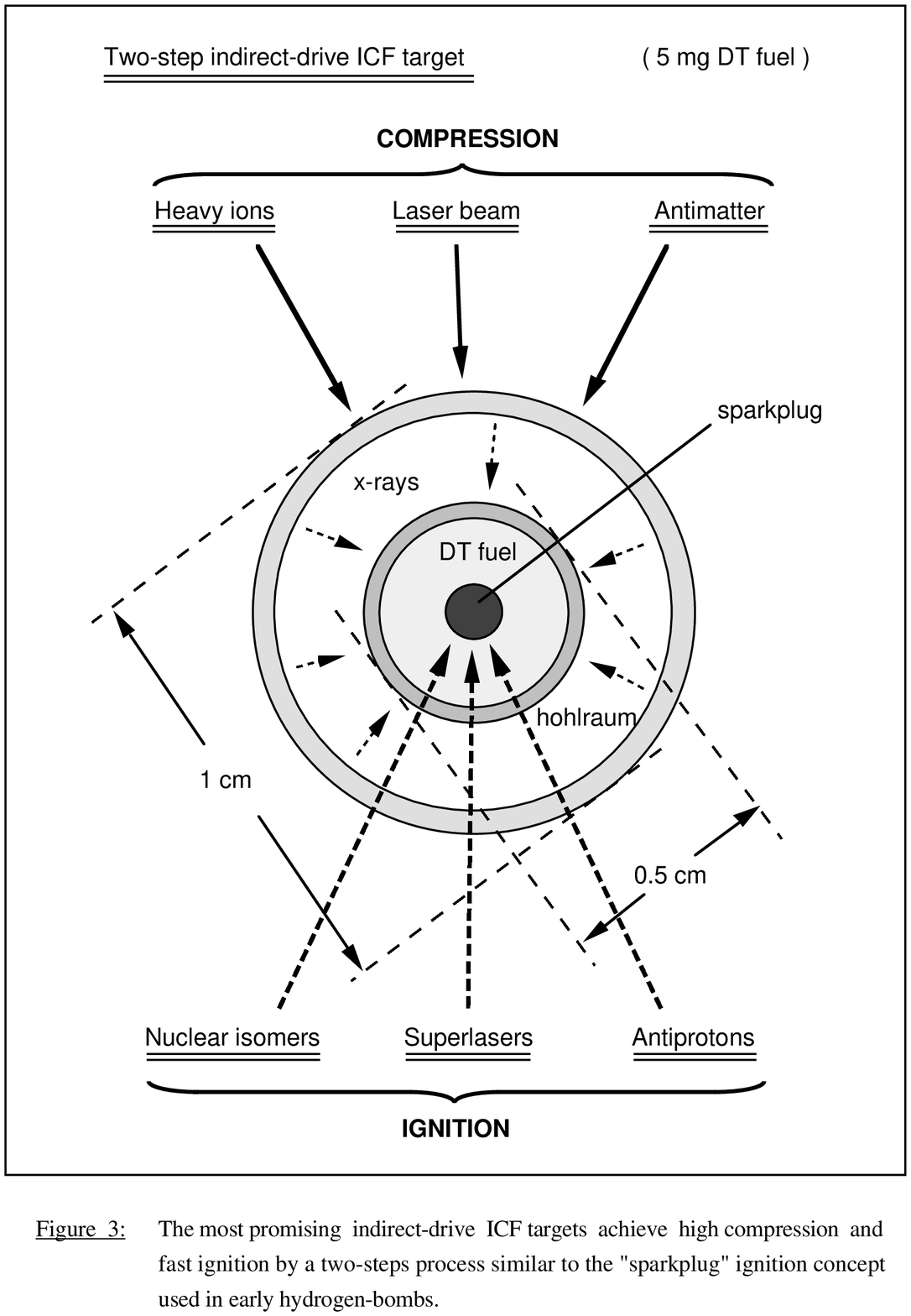}}
\end{center}
\caption{Fig.~25 of Ref.~\cite{GSPON1997A}}
\label{fig:ADV}
\end{figure}

   Compressing the pellet without excessively pre-heating it is called ``adiabatic compression.'' In most countries the currently preferred method for this is to use high-energy laser beams, like at the NIF and LMJ facilities, or heavy-ion beams like in Germany where a number of studies have been made on this option \cite{GSPON1997A}.  The principle is to strongly heat the radiation case (i.e., the walls of the capsule containing the pellet, see Figs.~\ref{fig:ICF} and~\ref{fig:ADV}) which then creates a hot X-ray bath that implodes the pellet. 

  As laser or heavy-ion beam drivers are very large, their use is restricted to experimental facilities,  or to possible future ICF energy-production reactors.  For weaponized ICF devices, much more compact techniques are required:  for example antimatter, as is suggested at the top of Fig.~\ref{fig:ADV}.  However, techniques less ``exotic'' than antimatter may also be used as compact drivers for compression: high-explosive driven magnetic-compression \cite{YOUNG1996-,CHERN1996-,GSPON1998-}, or else high-explosive driven chemical-lasers \cite{BASOV1992-}, for example.

   To ignite the pellet after compression, a different method is required because ignition has to be achieved on a much shorter time-scale than compression.  For example, while a 1~MJ energy laser pulse of a few nanosecond duration is suitable for compression, ignition may require a pulse of only 1~kJ energy, but of a duration several thousand times shorter, i.e., less than one picosecond.\footnote{Hence, the name ``fast ignition,'' or ``fast heating'' \cite{KODAM2002-}.}  Thus, while high-\emph{energy} lasers are needed for compression, high-\emph{power} lasers are required for ignition.

Laser suitable for fast ignition are called petawatt lasers --- or ``superlasers'' because their pulse duration is so short that they concentrate so much energy in so little time that they can produce nuclear reactions directly \cite{GSPON1997A, LEDIN2002-}.   These superlasers are based on a technique called ``chirped pulse amplification'' that was invented in the mid-1980s, and which enabled a one-million-fold increase of the power of table-top lasers.  kJ-energy superlasers are in existence or under construction in only a few countries: the USA, Japan, France, UK, and Germany.

\subsection{Nuclear isomers}
\label{Sec.3.4}

   Nuclear isomers are long-lived excited states of an atom's nucleus. Some nuclear isomers are naturally occurring, but most of them are produced in artificial nuclear reactions by colliding beams of nuclei and particles.  Nuclear isomers decay back to the ground state after some time, which can vary from picoseconds to billions of years, therefore releasing energy in the form of gamma radiation.

If a method could be found to release that energy instantaneously in a gamma-ray burst, rather then slowly and at random over time, one would have a new method for controlled high-energy storage and release.

   The possibility of practical applications of isomers, including for military purposes, has been highlighted in the past few years by the claim that a method for the controlled energy release from one particular isomer, Hafnium-178m, had been discovered.  However, more precise experiments recently showed that this is actually not the case \cite{BECKE2005-}.

   Consequently, the state of affairs with nuclear isomers is somewhat similar to that of ``super-heavy elements'' some decades ago, when the theoretical prospect for the large scale synthesis of these elements was uncertain: it is only recently that it has become definitely clear that the production cross-sections of stable super-heavy elements are too small for any practical applications to be feasible.

   In the case of nuclear isomers, the possibility of the existence of a suitable isomer for practical applications (such as a compact energy sources for compressing and/or igniting thermonuclear pellets) is entirely open.  It could be a matter of pure chance that such an isomer exists, and that a method for the controlled release of its energy could be found.

   Research on nuclear isomers is extensively carried out at a number of facilities around the world, especially in the USA, Russia, France, Japan, and Germany.  These facilities are based on high-intensity radioactive beams and are generally justified in view of their prospect for studying short-lived nuclear species, which are still not fully understood from the fundamental nuclear physics point of view.   But should an isomer suitable for practical applications exist, it is most likely that it would be found at one of these facilities.

   In this context it is important that Japan is building an ambitious project to generate the world's most powerful beams of protons, as well as a large radioactive-ion-beam facility \cite{CARTL2003-}; and that Europe is planning a next-generation radioactive-ion-beam facility called EURISOL to come online sometimes after 2013 \cite{CARTL2004-}.

\subsection{Antimatter}
\label{Sec.3.5}

Antimatter has the distinctive feature of having a flavor of being a ``science fiction technology'' while in fact there is no fundamental scientific problem with that technology and its potential applications \cite{ANTIM2005-}.  The problems are that antimatter is expensive to produce in large quantities at present, and that numerous engineering problems have still to be solved before it can be used in practical applications.

The situation is similar to that of microcomputers or hand-held cellular telephones: their scientific feasibility was established in the 1950s, but their construction and large-scale manufacturing required thirty to forty years of engineering research and development.

For example, the theory I presented at a 1986 conference, demonstrating the scientific feasibility of a new generation of nuclear weapons in which the few kilograms of fissile primary are replaced by microgram amounts of antihydrogen, i.e., \cite{GSPON1986A, GSPON1986B, GSPON1987-},  has never been disputed.  On the contrary, that presentation is now recognized as being the first presentation at an open scientific conference of the correct physical processes leading to the ignition of a thermonuclear explosion using less than a few micrograms of antimatter as trigger, and the correctness of the theory has been confirmed in many reports, see, e.g., Refs.~\cite{PERKI2004-, SHMAT2005-}. 

As a matter of fact, antimatter is seriously considered as a potential energy source for a new generation of nuclear explosives since at least the end of the 1940s, especially by Edward Teller and collaborators in the United States \cite{GSPON1986B}.  

Antimatter is currently a very active area of research, characterized by a considerable overlap with fundamental nuclear and sub-nuclear research because antiparticles are particularly powerful tools for studying the fundamental laws of physics at very high energies.  There is therefore a large consensus, as well as a convergence of interests, which make that the development of antimatter science and technology does not need to be funded by large defense programs, simply because it is a high priority item for pure scientific research as well.

Consequently, the number of antimatter-based research facilities has mushroomed around the world, with the result that some large and primarily national facilities (e.g., in Japan \cite{CERN-1997-} and Germany \cite{FEDER2002-, CARTL2003-}) are becoming directly competitive with intergovernemental laboratories (such as CERN) which should be the proper place for conducting ``pure science'' in a strictly international setting.  It is also remarquable that in the past ten years a rather exotic field like antimatter-plasma physics has become a large subject of research, with several percent of all professional advanced plasma-physics publications dedicated to it, including contributions from countries such as Brazil which has developed a strong theoretical basis for mastering the subject.

As a result, by 2010, the world's two largest antiproton factories will be in Europe, at CERN near Geneva, and GSI near Darmstadt, i.e., only about 1000 km apart from each other.  This has not escaped the attention of US scientists working on the practical applications of antimatter, who emphasized that
\begin{quote}
\emph{``a dedicated antiproton source (the main barrier to copious antihydrogen production) must be built in the US, perhaps as a joint NASA/DOE/NIH project. This is because the only practical sources in the world are at CERN and the proposed facility at GSI in Germany''}
\cite{NIET02004-}.
\end{quote}

This may look as a quite unusual situation: why would the world's richest and most powerful country not have a large dedicated antiproton source?  Part of the answer to this is already contained in a twenty-years old report of the RAND corporation which basically concluded that as long as US scientists would have full access to antimatter produced at CERN ``a production/accumulation facility, such as the one at CERN, although desirable, wouldn't in the near future have to be built in the United States'' \cite[p.~43]{AUGEN1985-}.  In this spirit, one could also add that it could certainly be in the interest of the United States to let Europe have a few unique advanced and costly facilities in order to motivate its researchers, and to attract those from Eastern Europe and Russia.\footnote{Similar reasons explain the support given by the United States to the creation of CERN in the early 1950s \cite{GSPON2004H}.}

However, the main reason is more probably that the production of antiprotons is not the most difficult technological problem in the way of practical applications of antimatter.  In fact, the $10^{14}$ antiprotons per year to be produced in 2010 at the AD facility at CERN, or at FAIR at GSI, correspond already to about 1~nanogram per year.  If presently available technology would be used to build a fully dedicated ``antimatter factory,'' rather than a general purpose ``research facility,'' one could easily produce more that 1~microgram of antimatter per year \emph{right now} \cite{GSPON1987-}.  As 1~microgram is sufficient to trigger one thermonuclear weapon, such a facility will only be a factor 365 away from the implicit goal that the US and Soviet governments set forth in the 1940s, namely to produce enough material for making one atomic bomb every day!

As a matter of fact, the United States and other countries are still investigating the best technique for producing very large quantities of antimatter.  One such technique is based on the idea of creating a ``quark-gluon plasma,'' which is studied in laboratories such as CERN, RHIC at BNL (the Brookhaven National Laboratory near New York), and FAIR at GSI.  Creating such a plasma is essentially trying to reproduce in the laboratory what happened at the beginning of the universe, a tiny fraction of a second after the big bang.  At that moment there were equal amounts of matter and antimatter in the universe, all matter and energy being in a so-called ``primordial plasma'' state. If the cooling-down process of the primordial plasma into either matter or antimatter could be controlled, one would possibly have the most efficient method for producing antimatter on a large scale!  There is therefore no surprise that weapons-laboratory scientists are in fact much interested by this supposedly purely ``astrophysical'' state of matter \cite{SOLTZ2003-}, especially since the idea itself originated from Edward Teller and collaborators in 1973 already \cite{CHAPL1973} (see also \cite{GSPON2003B}).

  There is also another serious and almost trivial question:  How can antimatter safely be stored and manipulated as it spontaneously interacts with any matter coming close to it?  This is the subject of considerable research for which many options are in competition.  Clearly, there is little reasons for building a full scale antimatter factory before this question is fully answered, and the related technologies sufficiently developed.

\subsection{Micro/nano-technology and nuclear bullets}
\label{Sec.3.6}

Possibly the most fascinating aspect of antimatter technology is that it leads to the prospect of ``nuclear bullets,'' i.e., thermonuclear explosive devices that would have the size of an egg and an explosive yield of a few \emph{tons} high-explosive equivalent.

In order for this to become possible, it is necessary to be able to store and deliver microgram amounts of antimatter in well controlled manners.  The most likely solution to this problem will probably require that antimatter be stored within a condensed material, in which special sites will function as micro- or nano-metric sized ``traps'' where particles of antimatter will be confined in a metastable state sufficiently away from ordinary matter.  Triggering the release of that antimatter could produce an explosion by interactions with the condensed material, or could be controlled, and the antimatter moved to another place if  it could momentarily behave as a superconducting fluid within the condensed material \cite[Appendix]{GSPON1986B}.\footnote{That idea actually originates from Dr.\ Jean-Pierre Hurni.}

\begin{figure}
\begin{center}
\resizebox{8cm}{!}{ \includegraphics{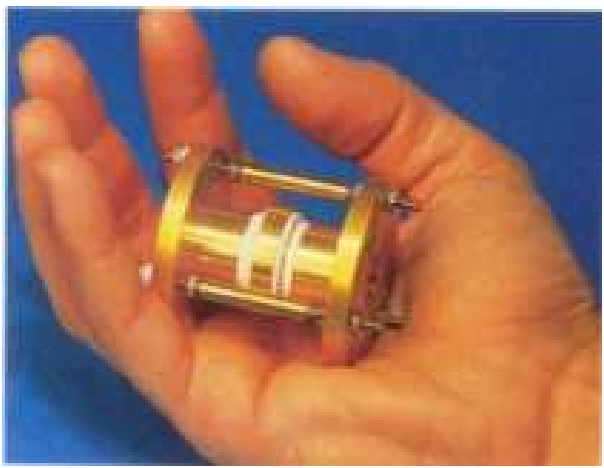}}
\end{center}
\caption[Antimatter bomb]{\emph{Small antimatter trap. (Dan Brown, Angels and Demons, Atria Books, 2005, ISBN 0-7432-7506-3, p.~71. Credit: CERN/ Photo Researcher, Inc.)}} 
\label{fig:anti-trap}
\end{figure}

Such a condensed-matter solution will be much more suitable and rugged than currently used electromagnetic traps (see Fig.~\ref{fig:anti-trap}), and in line with the contemporary emphasis on micro-electromechanical engineering and nanotechnology \cite{GSPON2002B}.  Moreover, as already mentioned in the Sec.~\ref{Sec.2.4}, the  development of new high-yield earth penetrating weapons is also depending upon substantial development in the these same technologies  \cite{GSPON2002B}: it is therefore not impossible that an antimatter-triggered earth penetrating weapon could be more robust than one fitted with a fission-primary --- which would certainly result in a weapon that would be radiologically much cleaner since no fissile material would be used at all.

In summary, if it were not for a few electro-mechanical devices that may be necessary for some essential tasks such as increasing the security, safety, and reliability of a weapon, or for triggering the initiating system, it is quite possible that future FGNWs could use no loose or moving parts at all (or else so much miniaturized that their ruggedness could be pushed to the limit).  In that case the weapons could be essentially compact solid-state devices, i.e., ``full like eggs,'' and therefore amenable to mass-production and scaling provided they would not use any process that is bound to a critical scale or mass.  This is clearly possible in theory with techniques such as inertial confinement fusion, antimatter, micro-electromechanical engineering, and nanotechnology.

\subsection{Pure antimatter bombs?}
\label{Sec.3.7} 

    During 2004 a number of newspaper articles referred to a US Air Force program related to development of new types of explosives and high-energy fuels based on antimatter, which emphasized the use of positron (i.e., anti-electron) rather than antiproton annihilation energy \cite{DAVID2004A, DAVID2004B}.  The difference with antiprotons is that positrons are 1876 times lighter, and therefore worth 1876 times less energy when annihilating.  However, since positrons are much easier to be produced than antiprotons, and could even be efficiently produced using superlasers, it could be that positrons may be competitive to antiprotons as an antimatter-based source of energy for certain applications.

    Another major difference between positrons and antiprotons is that upon annihilation positrons produce two 0.511~MeV gamma-rays, while antiprotons produce on average four 187~MeV gamma-rays, and three 236~MeV pions which are  directly suitable for heating a thermonuclear fuel.  Nevertheless, by letting them simply annihilate in any material, a small amount of positrons can produce quite a big bang: About 21.4~\emph{tons} for one milligram of positrons.

   It turns out that this latter possibility has been popularized by a highly successful novel, ``Angels and Demons,'' in which a terrorist group called the \emph{illuminati} is threatening to destroy the Vatican with 0.25~g of positrons stolen from CERN, i.e., an amount worth 5~\emph{kt} of explosive yield \cite{BROWN2000-}.  As a result, this novel has succeeded in getting millions of people to understand that a new generation of nuclear weapons was scientifically possible, therefore starting some kind of a public discussion on the prospect of such weapons.  The irony is that this is precisely one thing I have been trying to do in the past twenty years --- without success.  Moreover, there is only one CERN scientist mentioned by name in the novel: Nobel laureate Georges Charpak \cite[Sec.8]{BROWN2000-}, who was my first boss when I started working at CERN in 1971.  Unfortunately, when I later published a book in which I mentioned my worries about the prospect of antimatter weapons \cite{GRINE1984-}, Charpak did not like it at all: In a 1997 book about nuclear energy and nuclear weapons he called me an \emph{illuminati} because of these concerns \cite[p.16]{CHARP1997-}.  Somehow the author of ``Angels and Demons'' picked-up the ideas of antimatter-weapons and \emph{illuminati} --- and took them as the basis for a bestseller...

   Returning to the subject of positron-based antimatter explosives, it should be stressed that by taking a suitable material and letting a large number of positrons annihilate in it, that material could be brought to a sufficiently high temperature to become an effective source of soft X-rays that could be used to implode a secondary containing a thermonuclear pellet.  Therefore, just like with antiprotons, it makes much more sense using positrons as a primary charge (i.e., a trigger) rather than as the main charge of an explosive (See top of Fig.~\ref{fig:ICF} and~\ref{fig:ADV}).  As a matter of fact, this is similar to the economics of three-stage second generation nuclear weapons, where minimal amounts of expensive Pu-239 (or U-235) and tritium are used to burn a sizable amount of inexpensive $LiD$ and a lot of ``lower-grade'' uranium or plutonium:  in the case of fourth generation nuclear weapons, a minimal amount of very expensive antimatter or nuclear isomers will be used to trigger the burn of a larger amount of ``less-costly'' thermonuclear fuel such as $DT$ or $Li_2DT$.

   In conclusion, whether based on antiprotons or positrons, pure antimatter weap\-ons are unlikely to be manufactured on a large scale because of the high cost of both types of antimatters.  Moreover, in the case of positrons, high-density storage is less likely to be possible than with antiprotons.  This is because positrons annihilate with electrons, which implies that finding or designing an appropriate material in which they could be stably confined in large numbers is more difficult, something that is well known from over sixty years of research on the interactions of positrons with materials.

\section{Target coupling}
\label{Sec.4}

In this section we intend to examine the military effectiveness of FGNWs and to compare it to those of non-nuclear and previous generation nuclear weapons.  As before we will focus on the case of $DT$-fusion based FGNWs, and for the purpose of this section assume that such weapons can be built and deployed.

The basic concept used in quantifying and comparing the effectiveness of weapons is that of ``coupling,'' which refers to the efficiency of the way the energy\footnote{As will be seen in a few lines, because energy and momentum are separately conserved in physical interactions, one should in principle distinguish between ``energy coupling'' and ``momentum coupling.''  But since most of the damage comes from the energy that is inelastically transferred to a target, ``energy coupling'' provides a measure of the damage that can potentially be made.} of the weapons is transferred to a given target in order to damage or destroy it.  This implies that it is necessary to understand under which form energy is initially released by the weapon to the environment (Sec.~\ref{Sec.4.1}), and how this energy is transferred to the target (Sec.~\ref{Sec.4.2}).

  But before going into this, it is useful to recall a few elementary facts about physical interactions between a projectile and a target, which will illustrate the main concepts that will be used in the sequel of this section.

  Let us therefore consider a projectile of mass $m$ and initial velocity $v_0$, and a target of mass $M$ and velocity $V$ that is initially at rest, i.e., $V_0=0$.  One may think of the projectile as a bullet, but it could just as well be a neutron from a nuclear weapon, or else an air molecule pushed by a shock wave propagating through the atmosphere.

  To simplify we consider the one-dimensional case in which the projectile and the target are restricted to move along a line.  Because energy and momentum are separately conserved one has two equations:  One for the momentum
\begin{equation}\label{mom}
           m v_0 = m v + MV,
\end{equation}
where $v$ and $V$ are the final velocities of the projectile and the target, respectively;  And another one for the energy
\begin{equation}\label{ene}
           \frac{1}{2}m v_0^2 = \frac{1}{2}m v^2 + \frac{1}{2}M V^2 + E_{\text{int}},
\end{equation}
where apart from the final kinetic energies a term $E_{\text{int}}$ has been added to take into account the changes in target (and possibly projectile) internal energies.  There are therefore three unknowns, and only two equations.  However, in several important cases one can make reasonable hypotheses, and solve the problem.

   The first important case is when the interaction is \emph{elastic}, that is when $E_{\text{int}}=0$. This is a reasonable first approximation to the case where the projectile is a molecule of air put in motion by a shock wave, and the target a layer of a solid material hit be the shock wave.  One can than solve the system, and after some algebra find the energy $E_t$ given to the target as a function of the initial kinetic energy $E_p = \frac{1}{2}m v_0^2$ of the projectile.  The answer, which can be found in many elementary books of mechanics, is
\begin{equation}\label{ela}
           E_t = 4 \frac{mM}{(m+M)^2} E_p \approx 4 \frac{m}{M} E_p,
\end{equation}
where the approximation correspond to the limit $m \ll M$. Consequently,  when the projectile mass is much smaller than the target mass, and the interaction is elastic, very little energy can be transferred from the projectile to the target, and the coupling is very small. 

   The second important case is when the interaction is fully \emph{inelastic}, that is when the projectile entirely penetrates into the target, and all its energy and momentum are transferred to it.  As the final velocity of the projectile and target will be equal, $v=V$, one has from Eq.~\eqref{mom}
\begin{equation}\label{vel}
           V = \frac{m}{(m+M)} v_0,
\end{equation}
and solving for $E_{\text{int}}$ one finds
\begin{equation}\label{ine}
           E_{\text{int}} = \frac{M}{(m+M)} E_p \approx E_p,
\end{equation}
where the approximation correspond to the limit $m \ll M$. Consequently,  when the projectile mass is much smaller than the target mass, and the interaction is fully inelastic, almost all of the projectile energy is transferred to the target in the form of internal energy  --- which contrary to kinetic energy directly contribute to damaging the target.  This situation corresponds to the case of a high energy neutron penetrating deep into a target material, where after a number of elastic and inelastic interactions with individual nuclei, most of its energy ultimately ends up in internal energy, i.e., heating of the target.

\subsection{Initial energy from conventional or nuclear weapons}
\label{Sec.4.1}

   In non-nuclear weapons, as is well known, the lethal energy is initially in the form of heat and pressure, i.e., the two basic manifestations of thermodynamic energy.  This is because the chemical detonation reaction leads to a very fast decomposition of the explosive material, so that the reaction products left after the detonation are very hot and consequently under very high pressure.  The detonation products will therefore tend to push the directly adjacent materials away, launching a shock wave through all surrounding materials, which will warm up both under the effect of this shock wave and because of the heat radiated away by the hot detonation products.

In a nuclear weapon the situation is quite different because the lethal energy that affects the immediately surrounding materials is mostly in the form of a variety of radiations: \emph{electromagnetic radiations} such as X- and gamma-rays;  \emph{nuclear radiations} such as neutrons;  and possibly \emph{sub-nuclear radiations} such as pions if the weapon contained a significant amount of hadronic antimatter (i.e., antiprotons rather than positrons).

  Consequently, since the kinetic energy of the expanding materials of a nuclear bomb generally corresponds to a small fraction of the radiated energy, the immediate vicinity of a nuclear explosive is that of an extremely-intense pulsed-source of radiations. Depending on the type of the bomb, the dominant kinds of emitted radiations are as follows:

\begin{itemize}

\item \emph{Hot fission bomb:} soft X-rays and some fission neutrons;

\item \emph{H-bomb:} soft X-rays and some fission and fusion neutrons; 
 
\item \emph{Pure fusion bomb:} 14 MeV neutrons and soft X-rays;

\item \emph{Pure isomer bomb:}  0.1 to 5 MeV gamma-rays;

\item \emph{Pure positron bomb:} 0.511 MeV gamma-rays;

\item \emph{Pure antiproton bomb:} $\approx$~200 MeV pions and gamma-rays.

\end{itemize}

In all these devices it is assumed that the explosive charge is contained in  a casing such that any low-energy neutral or charged particles --- e.g., the X-rays from a fissioning material, or the $He^{4}$ nuclei produced in the $DT$ reaction~\eqref{DT-reaction} --- will not be able to escape, so that their energy will directly contribute to heating the bomb's materials and casing, which will thus radiate soft X-rays into the surroundings.

\subsection{Initial work from conventional or nuclear weapons}
\label{Sec.4.2}

As we have recalled at the beginning of the previous subsection, the initial work of a conventional explosive on the surrounding materials is primarily to launch a shock wave through them, and secondarily to warm them up through the effect of this shock wave and the heat radiated by the hot detonation products.

In the case of nuclear explosives the situation is more complicated because the different kinds of radiations can have a variety of effects, especially if they are very penetrating, as is the case for high-energy neutrons and gamma-rays.  The most important of these effects are as follows:

\begin{itemize}

\item \emph{Generate a fire-ball (in air or a material)}.  This is primarily the effect of the soft X-rays which have a relatively short mean-free path in any material, including air.  The material will heat up and the resulting fire-ball will radiate longer wavelength electromagnetic energy, i.e., a heat wave leading to various thermal effects.

\item \emph{Launch a shock-wave (in air or in a material)}.  This is primarily the result of the expansion of the soft X-rays-generated fire-ball into the surroundings, which launches a shock wave leading to blast effects.

\item \emph{Heat the surface of a material}.  Hard X-rays and low-energy gamma-rays able to propagate over some distances in low-density intervening materials (e.g., air) will be absorbed at the surface of any high-density material.

\item \emph{Ablate a material and produce a shock wave in it}.  If surface heating is sufficiently strong, the material will vaporize (i.e., ``ablate'') and by reaction (i.e., ``rocket effect'') a large pressure will be exerted on it, launching a shock-wave into the material.

\item  \emph{Accelerate or compress a material}.   If the ablation pressure is sufficiently strong, a material can be accelerated to high velocity by rocket-effect; and if the ablation pressure is simultaneously exerted on all sides, a material can be compressed to high-density as is the case of the secondary in a two-stage thermonuclear weapon.

\item \emph{Transfer momentum to a material}.  Either directly through the effect of radiations, or indirectly by means of shock waves propagating through an intervening medium, momentum can be transferred to a material which can be directly accelerated to high velocity without being ablated.

\item \emph{Heat the volume of a material}.  Penetrating high-energy radiations  (neutrons, pions,\footnote{Volume heating by pions is the process by which a very small amount amount of antiprotons can ignite a ``spark'' in a compressed thermonuclear fuel and thus trigger a large-scale thermonuclear detonation \cite{GSPON1986A}.} or high-energy gamma-rays) will easily cross a low-density intervening medium such as air and deposity their energy deep into any high-density material.  As a result, a substantial (i.e., centimeter to meter-thick) layer of a bomb-irradiated material can be brought to a temperature sufficiently high for it to melt, vaporize, or even explode.

\item \emph{Energize a working material}.  A special case of volume heating is that in which a ``working material'' is intentionally placed near a nuclear explosive in order to heat it to high-temperature so that it can do mechanical work on other materials.  This is the nuclear analog of a steam machine, in which super-heated water (i.e., steam) is used to produce motion.

\item \emph{Forge and project missiles}.  A super-heated working material can be used to forge a material into a missile and project it to a large distance.

\item \emph{Form and send high-velocity jets}. A super-heated working material can be used to form and send high-velocity (plasma) jets to some distance.

\end{itemize}

This list calls for three remarks:

\begin{enumerate}

\item The above list includes only the primarily ``mechanical'' and ``thermodynamical'' effects of nuclear explosives.  Important non-thermo-mechanical effects such as the production of an electromagne\-tic-pulse affecting electronic equipments, and the prompt or delayed radiations affecting living bodies (and electronic equipments to some extent), can be considered as collateral effects in that perspective.

\item As was stressed in the introduction to this section, many physical processes (such as  energy \emph{and} momentum transfer, transformation of kinetic into internal energy, etc.) have to be simultaneous taken into account, so that none of the effects in the list are ``pure effects'' that would be fully independent from the other effects.

\item Because they produce mainly blast and thermal effects, first and second generation nuclear weapons can basically be considered as gigantic conventional weapons --- except of course for their radioactive fallout and other nuclear-radiation effects.

\end{enumerate}

\subsection{Coupling to homogeneous and heterogeneous targets}
\label{Sec.4.3}

    From the previous discussion it is clear that a definition such as
\begin{center}
\begin{tabular}{|ll|}
\hline
   {\bf coupling : }
                 & {\it fraction of the total initial  explosion-energy}\\
                 & {\it actually transferred to an intended target,}\\
\hline
\end{tabular}
\end{center}
must be supplemented by due consideration to the physical processes involved in the energy transfer from the source to the target, and to the target response to the energy coupled to it.

This is because \emph{conventional explosives, and first and second generation nuclear explosives,} primarily couple their energy to the target by means of shock-waves propagating through an intervening medium: air, water, earth, rocks, etc. This means that the coupling of these weapons can be qualified as \emph{indirect}, independently on whether the target is (relatively) close or distant from the point of explosion.\footnote{It is only in outer-space nuclear explosions that direct coupling of X-rays is the main target interaction mechanism.}

In the case of \emph{fourth generation nuclear explosives}, however, the coupling can be qualified as \emph{direct}, unless the target is sufficiently far away from the point of explosion that the radiations are absorbed in the intervening medium before interacting with the target.  In other words, the fact that these weapons are primarily very intense sources of penetrating radiations means that they can produce direct work on the target, and therefore induce a very different response than if the target was just hit by a shock wave.

The reason is that when a shock wave strikes a high-density material after propagating in a lower density medium (e.g., striking the ground after propagating through air) most of the energy in the shock wave is reflected, and only a small fraction of the energy of the initial shock wave is given to secondary shock waves propagating through the target material.  Consequently, as is well known, indirect coupling by means of shock waves is very poor, because such waves are reflected at the boundaries between low and high-density materials.  For example, for both conventional and current generation nuclear weapons, less than 10\% of the energy striking a relatively heavy target (e.g., a main battle-tank, a bunker, or the ground) is actually coupled to it, even for explosions very close to the target, i.e., ``surface bursts.''  As a matter of fact, for ideal (absolutely rigid) materials, incoming shock waves are fully reflected.

For this reason, other methods of indirect coupling than shock waves had to be invented in order to increase target coupling.  This led to the development of   ``kinetic-energy penetrators'' and ``shaped-charges'' to defeat tank walls and reinforced concrete bunkers, and of ``earth-penetrators'' to defeat underground targets \cite{GSPON2003A}.

 To summarize, because of the importance of shock reflection at boundaries, one can classify typical military targets in essentially two categories --- homogeneous and heterogeneous:

\begin{itemize}

\item  \emph{Homogeneous targets:}  slab of steel, concrete wall, ground, etc.

      Most of the impinging shock wave energy will be reflected, and the reduced energy transmitted shock will propagate through the target until it reaches the back surface --- where part of its energy will be reflected again, and some of it transmitted into the following medium.  There will be  damage only if the material's response is inelastic, i.e., in order of occurrence as function of shock strength: rear-side spall, crushing, melting, or vaporizing.

\item  \emph{Heterogeneous targets:} multi-layer or composite tank-wall, well-designed bunker,  structure deeply buried under a geological cover, etc.

      Because of the number of boundaries and possibly intentionally large differences in densities of the different layers, the secondary shocks will be reflected many times, and their energy diffused and absorbed in such a way that only a small fraction of the initial energy will be transmitted behind a properly designed multilayer shield.  Again, there will be damage only if the materials's yield strengths are exceeded, which may be of limited consequences if it does not happen behind the shield.

\end{itemize}

\subsection{FGNW coupling}
\label{Sec.4.4}

Let us suppose that the yield from an idealized $DT$-based FGNW consists of about 20\% in soft X-rays and 80\% in 14~MeV neutrons.  Let us also take into account that relative to a surface at some distance from the point of explosion, 50\% of each of these radiations will flow forwards, and 50\% backwards.

If we suppose that this weapon has a yield in the range of a few \emph{tons}, and is detonated in air at a relatively short distance from a target, say a few meters, most of the forwards going X-rays will reach the target where they will heat the surface, which may melt or vaporize up to the point of launching a shock into it.  Because that shock is produced directly on the target, it will be much stronger that if it have produced indirectly by means of a shock wave propagating through air, as well as much stronger that if it would have been produced by the expanding fire-ball hitting the target. 

The main effect, however, will come from the neutrons.  Not just because they correspond to a circa five times larger source of energy, but because neutrons can easily penetrate inside any material where they can deposit their energy locally and produce volume heating of the material.  This means that the coupling can be very high, since there is little reflection in comparison to shock waves, and little losses in comparison to surface effects where part of the absorbed energy is back-radiated or lost as kinetic energy of the ablated material.

Of course, to calculate the distribution with depth of volume heating by 14~MeV neutrons requires some knowledge of nuclear physics.  However, this knowledge is readily available since the 1950s because it is directly related to the problem of calculating the heat generated by neutrons and gamma-rays in the shields of nuclear reactors \cite{ETHERI1955-, GLASS1981-}.  Nowadays, there are also many computer simulation programs which enable to make these calculations much more easily and precisely.

\begin{figure}
\begin{picture}(120,200)(0,0)
\linethickness{  0.00mm}
\put(0,0){\framebox(120,200){}}
\LARGE
\linethickness{  0.25mm}
\linethickness{  0.50mm}
\put(   20.00,   40.00){\line(  1,  0){  100.00}}
\put(  120.00,   40.00){\line(  0,  1){  160.00}}
\put(  120.00,  200.00){\line( -1,  0){  100.00}}
\put(   20.00,  200.00){\line(  0, -1){  160.00}}
\linethickness{  0.50mm}
\put(   20.00,   40.00){\line(  0,  1){    4.80}}
\put(   20.00,   30.40){\makebox(0,0){   0
}}
\put(   36.67,   40.00){\line(  0,  1){    4.80}}
\put(   36.67,   30.40){\makebox(0,0){   5
}}
\put(   53.33,   40.00){\line(  0,  1){    4.80}}
\put(   53.33,   30.40){\makebox(0,0){  10
}}
\put(   70.00,   40.00){\line(  0,  1){    4.80}}
\put(   70.00,   30.40){\makebox(0,0){  15
}}
\put(   86.67,   40.00){\line(  0,  1){    4.80}}
\put(   86.67,   30.40){\makebox(0,0){  20
}}
\put(  103.33,   40.00){\line(  0,  1){    4.80}}
\put(  103.33,   30.40){\makebox(0,0){  25
}}
\put(  120.00,   40.00){\line(  0,  1){    4.80}}
\put(  120.00,   30.40){\makebox(0,0){  30
}}
\put(  120.00,   40.00){\line( -1,  0){    4.80}}
\put(  120.00,   53.33){\line( -1,  0){    4.80}}
\put(  120.00,   66.67){\line( -1,  0){    4.80}}
\put(  120.00,   80.00){\line( -1,  0){    4.80}}
\put(  120.00,   93.33){\line( -1,  0){    4.80}}
\put(  120.00,  106.67){\line( -1,  0){    4.80}}
\put(  120.00,  120.00){\line( -1,  0){    4.80}}
\put(  120.00,  133.33){\line( -1,  0){    4.80}}
\put(  120.00,  146.67){\line( -1,  0){    4.80}}
\put(  120.00,  160.00){\line( -1,  0){    4.80}}
\put(  120.00,  173.33){\line( -1,  0){    4.80}}
\put(  120.00,  186.67){\line( -1,  0){    4.80}}
\put(  120.00,  200.00){\line( -1,  0){    4.80}}
\put(  120.00,  200.00){\line(  0, -1){    4.80}}
\put(  103.33,  200.00){\line(  0, -1){    4.80}}
\put(   86.67,  200.00){\line(  0, -1){    4.80}}
\put(   70.00,  200.00){\line(  0, -1){    4.80}}
\put(   53.33,  200.00){\line(  0, -1){    4.80}}
\put(   36.67,  200.00){\line(  0, -1){    4.80}}
\put(   20.00,  200.00){\line(  0, -1){    4.80}}
\put(   20.00,  200.00){\line(  1,  0){    4.80}}
\put(    8.75,  200.00){\makebox(0,0){    1200
}}
\put(   20.00,  186.67){\line(  1,  0){    4.80}}
\put(    8.75,  186.67){\makebox(0,0){    1100
}}
\put(   20.00,  173.33){\line(  1,  0){    4.80}}
\put(    8.75,  173.33){\makebox(0,0){    1000
}}
\put(   20.00,  160.00){\line(  1,  0){    4.80}}
\put(    8.75,  160.00){\makebox(0,0){     900
}}
\put(   20.00,  146.67){\line(  1,  0){    4.80}}
\put(    8.75,  146.67){\makebox(0,0){     800
}}
\put(   20.00,  133.33){\line(  1,  0){    4.80}}
\put(    8.75,  133.33){\makebox(0,0){     700
}}
\put(   20.00,  120.00){\line(  1,  0){    4.80}}
\put(    8.75,  120.00){\makebox(0,0){     600
}}
\put(   20.00,  106.67){\line(  1,  0){    4.80}}
\put(    8.75,  106.67){\makebox(0,0){     500
}}
\put(   20.00,   93.33){\line(  1,  0){    4.80}}
\put(    8.75,   93.33){\makebox(0,0){     400
}}
\put(   20.00,   80.00){\line(  1,  0){    4.80}}
\put(    8.75,   80.00){\makebox(0,0){     300
}}
\put(   20.00,   66.67){\line(  1,  0){    4.80}}
\put(    8.75,   66.67){\makebox(0,0){     200
}}
\put(   20.00,   53.33){\line(  1,  0){    4.80}}
\put(    8.75,   53.33){\makebox(0,0){     100
}}
\put(   20.00,   40.00){\line(  1,  0){    4.80}}
\put(    8.75,   40.00){\makebox(0,0){       0
}}
\linethickness{  0.75mm}
\qbezier[200](   20.00,  149.33)(   21.67,  156.00)(   23.33,  162.67)
\qbezier[200](   23.33,  162.67)(   25.00,  158.67)(   26.67,  154.67)
\qbezier[200](   26.67,  154.67)(   28.33,  146.67)(   30.00,  138.67)
\qbezier[200](   30.00,  138.67)(   31.67,  132.00)(   33.33,  125.33)
\qbezier[200](   33.33,  125.33)(   35.00,  120.00)(   36.67,  114.67)
\qbezier[200](   36.67,  114.67)(   38.33,  110.00)(   40.00,  105.33)
\qbezier[200](   40.00,  105.33)(   41.67,  102.00)(   43.33,   98.67)
\qbezier[200](   43.33,   98.67)(   45.00,   95.33)(   46.67,   92.00)
\qbezier[200](   46.67,   92.00)(   48.33,   90.00)(   50.00,   88.00)
\qbezier[200](   50.00,   88.00)(   51.67,   85.33)(   53.33,   82.67)
\qbezier[200](   53.33,   82.67)(   55.00,   80.33)(   56.67,   78.00)
\qbezier[200](   56.67,   78.00)(   58.33,   76.00)(   60.00,   74.00)
\qbezier[200](   60.00,   74.00)(   61.67,   72.33)(   63.33,   70.67)
\qbezier[200](   63.33,   70.67)(   65.00,   69.33)(   66.67,   68.00)
\qbezier[200](   66.67,   68.00)(   68.33,   66.33)(   70.00,   64.67)
\qbezier[200](   70.00,   64.67)(   71.67,   63.33)(   73.33,   62.00)
\qbezier[200](   73.33,   62.00)(   75.00,   61.00)(   76.67,   60.00)
\qbezier[200](   76.67,   60.00)(   78.33,   58.67)(   80.00,   57.33)
\qbezier[200](   80.00,   57.33)(   81.67,   56.67)(   83.33,   56.00)
\qbezier[200](   83.33,   56.00)(   85.00,   55.00)(   86.67,   54.00)
\qbezier[200](   86.67,   54.00)(   88.33,   53.33)(   90.00,   52.67)
\qbezier[200](   90.00,   52.67)(   91.67,   52.00)(   93.33,   51.33)
\qbezier[200](   93.33,   51.33)(   95.00,   51.00)(   96.67,   50.67)
\qbezier[200](   96.67,   50.67)(   98.33,   50.00)(  100.00,   49.33)
\qbezier[200](  100.00,   49.33)(  101.67,   48.67)(  103.33,   48.00)
\qbezier[200](  103.33,   48.00)(  105.00,   47.67)(  106.67,   47.33)
\qbezier[200](  106.67,   47.33)(  108.33,   47.00)(  110.00,   46.67)
\qbezier[200](  110.00,   46.67)(  111.67,   46.33)(  113.33,   46.00)
\qbezier[200](  113.33,   46.00)(  115.00,   45.67)(  116.67,   45.33)
\put(   70.00,  173.33){\makebox(0,0){CH$_2$ target
}}
\put(   70.00,   13.33){\makebox(0,0){Depth [cm]
}}
\put(  -13.33,  120.00){\makebox(0,0){\rotatebox{ 90}{Energy deposition [J/cm$^3$]
}}}
\end{picture}
\caption{\emph{Energy deposition by a 1~\emph{ton} equivalent point source of 14~MeV neutrons detonated 1~meter away from a thick slab of polyethylene. (Figure from Ref.~\cite{SAHIN2003-}.)}}
\label{fig:ch2}
\end{figure}
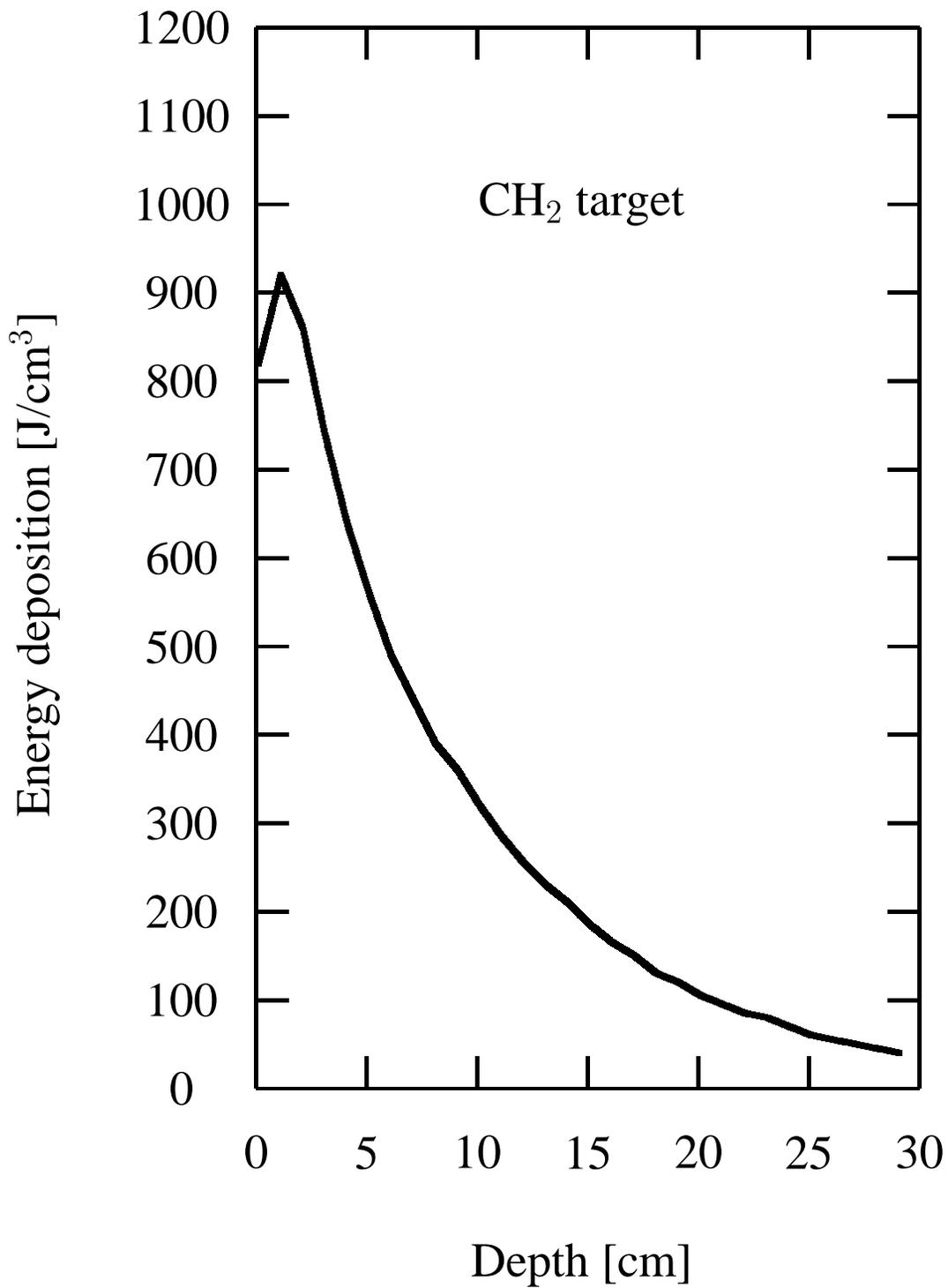

As an example, Fig.~\ref{fig:ch2} shows the neutron heating effect of a 1~\emph{ton} equivalent point source of 14~MeV neutrons detonated 1~meter away from a thick slab of polyethylene ($CH_2$), taken as representative (from the neutron-heating point of view) of the light materials used in modern multi-layered tank armor \cite{SAHIN2003-}.  As can be seen, heating is maximum at about 2~cm below the surface, and then decays exponentially with a half-length of about 10~cm.\footnote{If the heating of the surface by X-rays is included, the maximum would disappear and energy deposition would decrease with depth starting from the surface.}  Therefore, the energy deposited  in the first 10~cm has a density of about 0.5~kJ/cm$^3$, more than enough to vaporize the material.  Moreover, if the point of explosion is put at 30~cm rather than 100~cm, or if the explosive yield is increased from 1~to 10~\emph{tons}, the energy density would become comparable to that of the detonation products of a powerful chemical explosives. 

The same neutron heating calculation can be repeated with other materials:  earth, concrete,  aluminum, iron, uranium, etc.  The result is that the magnitudes, as well as the distributions with depth, are generally rather similar to those of light-weight materials such as $CH_2$, despite that in heavier materials the nuclear interactions of neutrons are very different from those in light-weight materials (i.e., much less elastic scattering, but more inelastic scattering instead).  It is only for very heavy material, or in materials such as uranium where 14~MeV neutrons can induce fission, that the magnitude of energy deposition can be larger by a factor of two or more.

  To summarize, and to phrase the results in a simplified form because what matters here are orders of magnitude rather than high precision, one has found that:

\begin{itemize}

\item Because most of the energy of a $DT$-based FGNW is in the form of highly penetrating neutrons, almost all of the forwards going energy is coupled into any target located less than a few meters away from the point of detonation.  This implies a coupling coefficient of almost 50\%, that is ten times higher than for any conventional or previous generation nuclear weapons;\footnote{The limit of 50\% of total FGNW energy coupling to a target can of course only be reached for a very wide target, e.g., the ground.  Nevertheless, for a finite width target, what matters is the factor of about ten in coupling increase relative to conventional or previous generation nuclear weapons.}

\item  The combined surface and volume heating effects of a 1~\emph{ton} FGNW detonated 1~meter away from any solid target leads to an energy deposition of about 1~kJ/cm$^3$ in the first 10~cm of any material.  

\end{itemize}

To make some further simplifications, this means that the energy deposition by 14~MeV neutrons is comparable to that of myriads of ``femto'' kinetic-energy or shaped-charge penetrators, and that while a 1 ton chemical explosion 1~m away from a 10~cm thick steel plate will barely damage it, a 1~\emph{ton} FGNW explosion at the same 1~m distance will burn a 1~m$^2$ hole through it.  Consequently, as suggested by Edward Teller in 1991, FGNWs with explosive yields of a \emph{few tons} may render many modern weapons obsolete:

\begin{quote}
\emph{``Shall one combine the newly acquired accuracy with smaller nuclear weapons (perhaps even of yields of a few tons) to be used against modern weapons such as tanks and submarines''} \cite{TELLE1991-}?

\end{quote}

\section{Thermonuclear-driven jets and projectiles}
\label{Sec.5}

While the ten-fold greater target-coupling of FGNWs is the first major difference between these weapons and those of the first and second generations, the second major difference is that the radiations from FGNWs can easily be used to energize a working material and therefore to produce other effects than just heat or blast.  As a matter of fact, this second difference was already one of the militarily-attractive feature of third generation nuclear weapons, where radiations from advanced nuclear or thermonuclear explosives were used in directed energy warheads (DEW) to produce powerful plasma-jets or X-ray laser-beams.

An implicit reason for considering such applications is that while about 50\% of the radiations moving forwards will directly interact with the target, the other 50\% will be scattered through air unless somehow used.   In other words, if a sizable fraction of the  backwards-radiated energy could be used to drive a forwards going device, the coupling and hence the lethality of FGNWs could be substantially increased.  Moreover, as was already considered in the case of third generation nuclear weapons, the same technique could be used for driving several such devices at once, possibly pointing in several directions, so that several targets could be simultaneously engaged by the same FGNW.

In this section we are therefore going to examing how FGNWs can be used to drive jets and projectiles --- leaving aside other possible applications such as driving powerful laser beams or electromagnetic-pulse generators.

But, just like at the beginning of Sec.~\ref{Sec.4}, we have to first recall that many different physical processes are at work when something is irradiated by an intense source of X-rays and neutrons.  For instance, several distinct processes can lead to the acceleration of a projectile, and depending on its physical composition and geometry, one or another of these processes will provide the dominant thrust mechanism accelerating the material to high velocity.

  For example, projectiles can be accelerated to very high velocities by means of X-ray ablation \cite{CAUBL1994-}, or by means of neutrons from a nuclear explosion \cite{RAGAN1977-, AVROR1993-, TRUNI1994-}.

As a matter of fact, the maximum momentum that the neutrons from a fourth-generation nuclear explosive can transfer to a projectile (and therefore the maximum velocity it may reach) can easily be estimated using Eq.~\eqref{vel}.  Indeed, each incoming neutron absorbed by the projectile will transfer a small amount of momentum to it, and for $N_n$ neutrons the total transferred momentum will be $N_n \times m_n \times V$, where $V$ is given by Eq.~\eqref{vel}.  As a neutron has a mass $m_n = 1.7 \times 10^{-27}$~kg, and thus a velocity $v_n = 52,000$~km/s for an energy of 14~MeV, the maximum velocity that can be reached by a projectile of mass $M$ is then
\begin{equation}\label{max}
           V_{\text{max}} = N_n(Y) \frac{m_n}{(m_n+M)} v_n
  \approx 130\frac{Y ~ \text{[\emph{tons}]}}{M ~ \text{[kg]}} ~~ \text{[km/s]},
\end{equation}
where $N_n(Y) = 1.45 \times 10^{21} Y$ is the number of neutrons corresponding to a yield of $Y$ [\emph{tons}] absorbed by the projectile.  If $M = 100$~kg and $Y = 1$~\emph{ton}, the maximum velocity is therefore about $1.3$~km/s.  However, as most of the energy absorbed by the projectile will contribute to heating rather than accelerating it, the projectile may not go very far before disintegrating --- unless only the rear part of the projectile is heated, and therefore acts as a working fluid pushing the front part ahead by rocket-effect.

Thus, as can be seen from this example, the practical design of thermonuclear-driven devices is most likely to be very complicated, so that the  discussion presented in the following subsection has to be understood as a description of idealized concepts, which may look quite different in practice.

\subsection{Conventional shaped-charges}
\label{Sec.5.1}

Shaped-charges of conventional explosives were first used on a large scale in warfare towards the end of World War Two when the Americans introduced the ``Bazooka,'' and the Germans the ``Panzerfaust'' anti-tank weapons.  The basic idea of these weapons was to overcome the poor coupling of the blast-waves from a high-explosive charge by converting a fraction of its detonation energy into the kinetic energy of a forward going jet of metal, which by exerting a much greater pressure at the point of impact could penetrate into thick armor steel.

   The actual concept was to use a hollow-charge of explosive lined with a thin sheet of metal (usually of conical shape) which upon detonation of the explosive would be squeezed towards the axis and vigorously projected forward in the form of a powerful jet. The details of the designs and performances of these weapons are given in many papers and books, but for a good introduction the original papers published in the 1940s are still up to date, e.g., \cite{BIRKO1948-}.

Of course, since these early days, the design of shaped-charges has considerably evolved \cite{DELON1995-}, and current weapons are perfected using very advanced computer simulations techniques. This enables to very precisely follow the collapse of the liner and the formation of the jet, as well as to optimize its performance in defeating various targets \cite{FAN--2001-}.  It is also by a sophisticated interplay between simulation and experiment that some of the most powerful shaped-charges have been designed and shown to be able to pierce up to 340~cm of high-strength armor-steel \cite{BAUM-1998-}.  In that case a large hemispherical charge lined with molybdenum was used (See Fig.~\ref{fig:Baum}), and remakable agreement between simulation and experiment was found.

 A final important aspect of shaped-charges is that they can be used not just to form and project a thin high-velocity jet, but also to forge and accelerate a fragment of substantially larger mass, albeit of lower velocity.  This is of interest because high-velocity jets tend to be unstable and therefore to have a rather short range in air, while forged fragments can travel much larger distances before striking a target \cite{DELON1995-}.

\subsection{``Nuclear'' and ``thermonuclear'' shaped-charges}
\label{Sec.5.2}

When addressing the question of nuclear- or thermonuclear-driven shaped-charges, it is important to first dissipate a possible misunderstanding due to the fact that in the 1950s the name ``atomic hollow-charge'' has been used for a very different concept, namely the possibility of using \emph{conventional-explosive}-driven shaped-charge jets to compress and ignite a thermonuclear fusion fuel \cite{HAJEK1955-}.  At that time this concept was thought to be plausible because jets of beryllium with velocities up to 90~km/s had been formed in some experiments.  But with hindsight it is now clear that any combination of such jets cannot be used to ignite a thermonuclear fuel because their kinetic energy is far too low to focus sufficient energy in a sufficiently short time on a pellet.

    The principle of a \emph{nuclear-driven}-driven shaped-charge is to replace the energy from the conventional explosive by the energy from a nuclear explosive.  Since most of the energy from a fission explosive is in the form of X-rays, this means using these X-rays to collapse a liner and form a jet, i.e., to replace the high-explosives's detonation products by a high-energy photon gas.  In theory, this concept is perfectly sound, and can be seen as a simple variation of the Teller-Ulam principle used to compress the secondary of a H-bomb. In practice, it seems that this idea was studied in the context of third generation nuclear weapons, but it is not known whether it was fruitful or not.  In particular, it could be that since there is no ``energy-multiplication'' as in a H-bomb secondary, the energy imparted to the jet was found to be too small to be of any use compared to the vast energy wasted in the fission explosive needed to drive the jet.

   Another concept which has been proposed is to replace the chemical explosive by a thermonuclear explosive.  This is because a thermonuclear detonation reaction is analogous to a chemical detonation reaction in the sense that they can both support a detonation-wave propagating through the explosives.  However, in comparison to chemical ones, thermonuclear detonation-waves are extremely thick in non-compressed fusion fuels \cite{FULLE1968-}.  Nevertheless, provided the thermonuclear fuel could be somehow precompressed, and properly ignited, it may be possible to conceive what could be called a ``thermonuclear shaped-charge'' \cite[p.~121]{WINTE1981-}.  The trouble with this concept is that the resulting device could be gigantic, with yields in the \emph{kt} to \emph{Mt} range, or more...

\subsection{FGNW-driven jets and projectiles}
\label{Sec.5.3} 

Merging and generalizing the ideas developed in Sec.~\ref{Sec.4} and in the previous subsections, it is obvious that both the X-rays and the neutrons from a FGNW explosive can be used  to heat and shock various working materials or fluids in order to achieve several purposes.  For instance:
\begin{enumerate}

   \item The neutrons and/or X-rays can drive or forge one or several projectiles or fragments (thermo\-nuclear-driven  ``grenades'' or ``guns'').  This can be achieved, for example, by absorbing the X-rays in a high density material which by rocket-effect will be accelerated, or by absorbing the neutrons in a working fluid which will forge and project a missile.

   \item The neutrons and X-rays can heat a working fluid and form jets (thermo\-nuclear-driven ``jets'').  In that case a working fluid super-heated by the neutrons will replace the high-explosive detonation products to pinch a liner and form a high-velocity jet. 

   \item The \emph{forwards} and/or \emph{backwards} X-rays and/or neutrons can separately or cooperatively drive the components of a multi-warhead weapon.

\end{enumerate}

\begin{figure}
\begin{center}
\resizebox{8cm}{!}{ \includegraphics{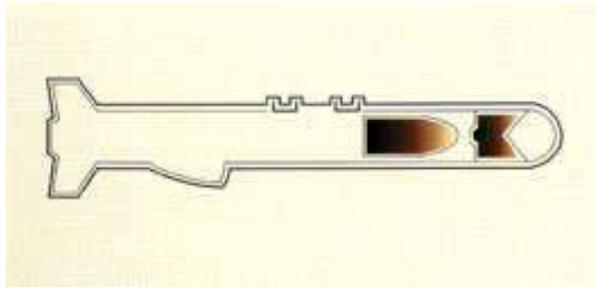}}
\end{center}
\caption{\emph{A single FGNW charge can drive several warheads at once:  Here, symbolically depicted, a forged-penetrator following a shaped-charge jet.}}
\label{fig:MWS}
\end{figure}

In the first two cases it is evident that a single FGNW explosive can project several missile or drive several shaped-charges at once, resulting in a multi-warhead system  (See Fig.~\ref{fig:MWS}).  In the third case, reference is made to the possibility of using the radiation moving backwards relative to a target as a driver for one or several forwards moving jets or missiles.

\begin{figure}
\begin{center}
\resizebox{8cm}{!}{ \includegraphics{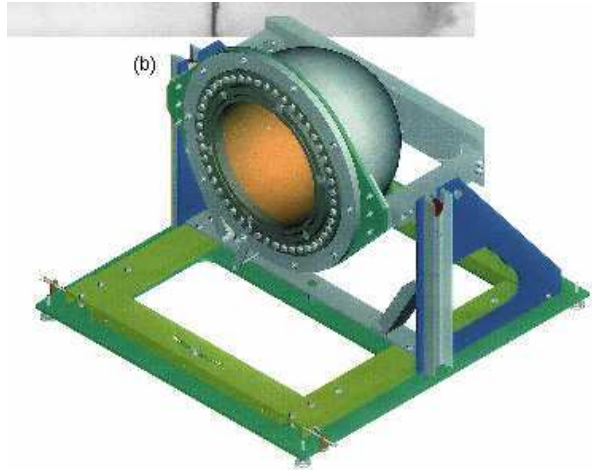}}
\end{center}
\caption{\emph{FGNW-driven shaped-charge concept: The hemispherical shell of high-explosives is replaced by a working-material such as polyethylene. An intense point-source of neutrons located at the focal point of the hemisphere is very uniformly super-heating this working-material, which will function just like the detonation-products of a super-high-explosive in collapsing the liner and driving the shaped-charge jet. (Illustration from Ref.~\cite{BAUM-1998-}.)} }
\label{fig:Baum}
\end{figure}

To give an example of a possible design of a dual-warhead system, one can refer to Fig.~\ref{fig:Baum} and imagine that the hemispherical shell of high-explosives is replaced by a simple working-material such as polyethylene.  If we now suppose that we have a very intense point source of neutrons at the focal point of the hemisphere, these neutrons will be able to super-heat the working-material which will function just as effectively as high-explosive detonation-products in driving the shaped-charge jet.  Consequently, the target (which is located on the left of Fig.~\ref{fig:Baum}) will be first irradiated by the forwards going neutrons, and then penetrated by the powerful jet driven by the backwards going neutrons.  Of course, if the point source of neutrons is replaced by a realistic fourth-generation nuclear explosive, the nascent jet will have to traverse its debris in order to move towards the target.  However, there are several ways to circumvent this problem, and the example given here is just intended to present the general idea.\footnote{While it is premature to go into many details, one can nevertheless remark that whereas the working material can be simple and inexpensive, it still has to be carried with the weapon.  It could therefore be
advantageous to make use of the delivery system's own material.  For example, the shaped-charge liner could be the fuel tank of a cruise-missile, and the working material the remaining of the fuel.}

    In order to develop such complicated fourth-generation warheads, it is clear that both sophisticated computer simulations and numerous experiments at inertial confinement fusion facilities such as NIF and LMJ will be necessary.  It is only after such simulations and experiments, which will correspond to neutron yields equivalent to at most a few \emph{kg} of high-explosives, that it will be possible to envisage doing experiments at neutron yields equivalent to a few \emph{tons}.  For this to happen much progress will have to be made, and it may turn out that driving jets and projectiles with FGNWs could be not much more effective than with third generation nuclear weapons. On the other hand, the large coupling efficiency of FGNWs discussed in the previous section will remain, and still be one the most important elements responsible for the military effectiveness of FGNWs.

\section{Collateral effects}
\label{Sec.6}
\begin{figure}
\begin{center}
\resizebox{8cm}{!}{ \includegraphics{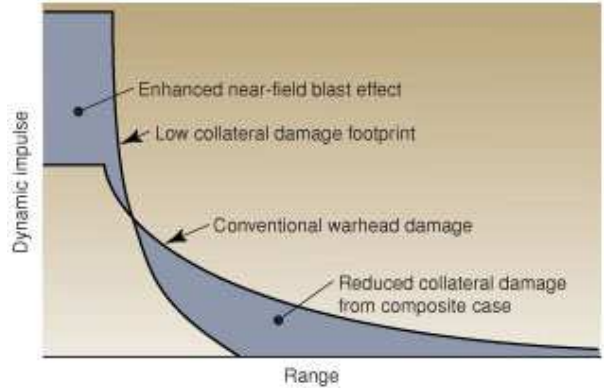}}
\end{center}
\caption{\emph{As a consequence of increased accuracy, new  weapons are required that will ``reduce destruction outside the radius of an intended target --- while enhancing destructive force on the target.'' (Illustration for Ref.~\cite{MURPH2003-}.)}}
\label{fig:Murphy}
\end{figure}

   As a consequence of the accuracy of the latest-generation of precision-guided missiles, it has become more and more important to reduce collateral damage, that is to ``reduce destruction outside the radius of an intended target --- while enhancing destructive force on the target'' \cite{MURPH2003-}, see Fig.~\ref{fig:Murphy}.\footnote{However, ``effective utilization of precision munitions demands that a premium be placed on the collection and analysis of target information'' \cite[p.~15]{YOUNG2000-}.}  In this section we try to see to what extent the characteristics of FGNWs correspond to this goal, especially in comparison to other types of weapons.

\subsection{Mechanical and thermal effects}
\label{Sec.6.1}

  The mechanical and thermal effects of conventional and nuclear weapons are well-known \cite{GLASS1977-, SANDM1982-}.  For instance, their scaling laws with explosive yield are simple power laws: direct proportionally $(\propto Y^1)$ for thermal effects, and third-root dependence $(\propto Y^{1/3})$ for blast over-pressure.  The factor of three difference in the exponent of these  power-laws makes that, in comparison to blast effects, thermal effects are generally negligible in conventional explosives, but dominant in \emph{Mt}-yield nuclear explosives --- which are in fact gigantic incendiary bombs \cite{HILL-1961-}.  This means that for \emph{kt}-yield nuclear weapons, and FGNWs with yields between 1 and 100~\emph{tons}, both effects should be taken into consideration.

   A first significant difference between $DT$-based FGNWs and all other types of explosives is that up to 80\%  of the yield is in form of high-energy neutrons, so that only about 20\% of the total yield contribute directly to heat and blast effects.  With proper scaling, this factor of 5 difference means that a FGNW will have a factor of 5 smaller incendiary effect, and a factor $\sqrt[3]{5}=1.7$ reduced blast effect\footnote{In reality, and for multi-warhead systems, the factor of 5 could be more like 3 or 4, leading to a smaller reduction in collateral heat and blast effects.}  --- provided on assumes that the energy of the neutrons will be absorbed either in the intended target, or else in a large volume of air that will not be sufficiently heated to significantly contribute to the heat and blast waves.  One can therefore conclude that for a given total yield, FGNWs will have somewhat reduced collateral effects in terms of heat and blast.

   The second significative difference between $DT$-based FGNWs and all other types of explosives is the high direct-coupling with intended targets made possible by the flux of high-energy neutrons.  Since according to the discussion in Sec.~\ref{Sec.4.4} this increased coupling corresponds to a factor of about 10 relative to a conventional or nuclear explosive, the factor of 5 considered in the previous paragraph is actually equivalent to a factor of about 50 when the comparison is not made in terms of total explosive yields, but in terms of the energy that is actually coupled to the intended target.  Under these conditions, the reduction factors in collateral heat and blast effects become truly significant, i.e., 50 and $\sqrt[3]{50}=3.7$, respectively.

  However, direct-coupling to a finite-size target has a $1/r^2$ dependence on the distance $r$ between the point of explosion and the surface of the target, and this distance should be on the order of a few meters at most for a circa 1~\emph{ton} FGNW to be effective.   This requires truly high accuracy in delivery, and a corresponding accuracy in the knowledge of the target coordinates.

Finally, as with all types of explosive weapons, debris will be sent at random to large distances from the target.  But since the kinetic energy available for sending these debris is directly related to blast energy, this collateral effect should be proportionally reduced in FGNWs.

\subsection{Prompt radiation effects}
\label{Sec.6.2}

  The lethal effect of high fluxes of high-energy neutrons has been much debated during the ``neutron bomb'' controversy of the 1970s to the mid 1980s \cite{SANDM1982-,SAHIN1985-}.  From these publications one knows fairly accurately the radiation dose produced by a pure $DT$-fusion explosion in air, which as a function of total yield $Y\text{~[\emph{tons}]}$ and distance $r\text{~[meter]}$ can be interpolated by
\begin{equation}
  D(r) = 10^8 \frac{Y}{r^2} \exp(-\frac{r}{360}) ~~~ \text{~~[rad]}.
\end{equation}
In this expression the radiation dose $D(r)$ is expressed in units of ``rad,'' rather than of ``rem'' or ``Gy,'' because for prompt radiation effects one does not need to take the relative radiobiological effectiveness into account.  

One can therefore find the distance below which the ``instant permanent incapacitation'' is close or equal to 100\%~:
\begin{center}
\begin{tabular}{|ll|}
\hline
       {\bf 1~\emph{ton} FGNW : ~~}   &  more than 10'000 rad below 100 m,\\
                               &  $>24~ ^{\text{o}}$C body temperature rise,\\
                               &  $>99$\% lethal within 1 hour.\\
\hline
\end{tabular}
\end{center}

\noindent And the distance beyond which the probability of survival is higher than 50\%~:
\begin{center}
\begin{tabular}{|ll|}
\hline
        {\bf 1~\emph{ton} FGNW : ~~}  &  less than 300 rad beyond 300 m,\\
                               & $ 1~ ^{\text{o}}$C body temperature rise,\\
                               & $<50$\% lethal within 1 month.\\
\hline
\end{tabular}
\end{center}

In these two boxes, the instantaneous full-body temperature rise produced by the given dose is calculate in order to provide an intuitive explanation for the prompt biological effect of high-doses of radiations.  As can easily be understood, an instantaneous full-body temperature rise from 37 to about $60 ^{\text{o}}$C will have a very big impact on physiology, which explains the immediate loss of consciousness and nearly instantaneous death.  On the other hand, a $1 ^{\text{o}}$C temperature rise will not have such a strong physiological effect, and death will be due to radiation sickness, which can be medically treated to some extent.

If instead of unprotected soldiers or people one considers heavy battle-tanks, which provided some shielding against radiations, the kill radius will be smaller \cite{SAHIN1985-}. Nevertheless, assuming a protection factor of 20, the tank crew will be instantly killed by a 1~\emph{ton} FGNW exploding at a distance of less than 20~meters.  In tank warfare language, this corresponds to a ``radiation-kill,'' while a direct hit would have produced a ``hard-kill.''

  In this context, a June 1994 interview of Russian Nuclear Energy Minister Viktor Nikitovich Mikhailov is significant.\footnote{V.N. Mikhailov is one of the scientists who helped develop the current generation of Russian nuclear weapons.}  According to a Reuter press release he said:
\begin{quote}
\emph{``a new generation of nuclear weapons could be developed by the year 2000 unless military research is stopped. [... This] fourth generation of nuclear weapons could be directed more accurately than current arms. [... The] new weapons could be programmed to wipe out people while leaving buildings standing. [... It is] a toss-up whether Russia or the United States would be the first country to devise the new arms''} \cite{MIKHA1994-}.   
\end{quote}

In conclusion, it is clear that FGNWs have a strong potential to be used as anti-personnel weapons, which in the case of the neutron bomb debate of the 1980s created a lot of controversy and even political unrest.  The corresponding debate with FGNWs is likely to be controversial as well, even though the absence of a fission-trigger in the weapon will make it more ``acceptable'' to some people.  Further complications will arise since neutrons can be rather effective against biological agents, and even against possible future ``nanorobots'' which could be designed to selectively disperse chemical and biological weapons.

\subsection{Delayed radiological effects}
\label{Sec.6.3}

In assessing the delayed radiological effects of FGNWs, a somewhat surprising finding was made in trying to compare the long-term radiological burden generated by their battle-field use to other forms of radioactive pollution \cite{GSPON2002A}.

The reason is that the two main delayed radiological effects of FGNWs (i.e., the dispersal of the unburnt tritium, and the activation of the ground and other materials by high-energy neutron irradiation) are relatively simple to estimate, which enables to calculate the total ``burden'' due to the expenditure of a given yield by integrating these estimates over time.  The same procedure can be applied to other weapons producing a different form of radiological pollution, and comparisons can be made.

Taking for example the use of depleted-uranium\footnote{It should be recalled that despite being a relatively benign radioactive substance (just like tritium), all grades of uranium are ``nuclear materials'' subject to special regulations according to national and international law.  The use of depleted uranium projectiles in the 1991 Gulf War therefore broke the 46 year long taboo against the intentional battle-field use of radioactive materials, which lasted from 1945 to 1991 \cite{GSPON2002A, GSPON2003-}.} penetrators as anti-tank weapons, and assuming comparable battle- and post-battle circumstancies (e.g, total dispersal of all radioactive materials, no cleaning-up of the battle-field, etc.), one obtains the following equivalence in terms of radiological burdens:
     \begin{center}
     \fbox{{\bf 1~\emph{ton}~~ FGNW} ~~ $ ~~ \approx ~~ $ ~~ 1~kg~~ U-238 }\\
     \end{center}

If one applies this equivalence relation to the 400 tons of U-238-munition which have been expended during the 1991 war in Iraq, one finds that the corresponding long-term radiological burden is equivalent to the hypothetical use of hundreds of thousands of FGNWs, each with a yield of 1~\emph{ton}.

Of course, both the methodology and the interpretation of the result can be criticized, especially if one focuses on short rather then long term effects.\footnote{As a matter of fact, reference~\cite{GSPON2002A} is a simple preliminary study: It would be important that a more comprehensive study be undertaken and published.} Indeed, there is a vast difference in the life-time of tritium and uranium, which is only partially compensated by the greater radio-toxicity of uranium relative to tritium.

Nevertheless, the comparison is useful to highlight the considerably smaller radioactive burden induced by FGNWs relative to the previous generations of nuclear weapons.  It can also be inferred that \cite{GSPON2002A, GSPON2003-}:

\begin{itemize}

\item Tritium dispersal and induced ground-radioactivity will to a large extent not impair further military action;

\item Just as it was the case with the use of depleted-uranium weapons, it will be possible for the proponents of FGNWs to argue that the radiological burden due to their use could be in some way tolerable;

\item  Many political leaders and large fractions of the public opinion may not object to the long term radiological impact of FGNWs;

\item In any case, with a tritium content of about 15 mg per \emph{ton} explosive equivalent, there will be 15~kg of tritium in an arsenal equivalent to one million 1-\emph{ton}-FGNWs,\footnote{A FGNW arsenal equivalent to 1~\emph{Mt} total yield is not unreasonable considering that a number of weapons could have maximum yields of about 100~\emph{tons}, and some up to about 1~\emph{kt}.} that is about the same tritium inventory as in one single full-size thermonuclear reactor \cite{GSPON2004A}.  Acceptance of civilian fusion power will therefore be linked to that of FGNWS.

\end{itemize}

\subsection{Electromagnetic effects}
\label{Sec.6.4}

All explosive processes, chemical or nuclear, produce electromagnetic effects of varying duration and intensity \cite{GLASS1977-}.  The best known of these is the ``electromagnetic pulse (EMP) effect'' which is most pronounced for fission explosions in the atmosphere where high-energy gamma rays generate a current of Compton recoil electrons, which produce further ionization so that the air becomes conductive \cite{KOMPA1959-, LONGM1978-}. But there other electromagnetic effects, especially associated with high-altitude fission explosions, such as the formation of man-made radiation belts around the Earth \cite{RABIN2003-}.  While in the first case the dominant source of the effect are gamma rays from excited or activated nuclei and the decay of fission products, in the second case it is the electrons and ions produced in these interactions and decays.  This means that fission is the dominant source of electromagnetic effects, so that pure fusion FGNWs will have significantly reduced effects of this type --- unless they are specially designed to use the fusion neutrons to drive an EMP generator, or to interact with a special material to produce a tailored effect.

\section{Conclusion}
\label{Sec.7}

It is difficult to write a concise conclusion for this essentially scientific paper because the military, technical, and political implications of FGNWs are so far reaching that it would take three other papers just to discuss each of these aspects with a minimum of rigor.

For this reason, instead of a true conclusion, there will be a list of ``theses'' which are meant to summarize in a few words some of the most important points that should be discussed in a comprehensive assessment of the prospect and implications of FGNWs.

\subsection{Military aspects}
\label{Sec.7.1}

\begin{enumerate}
\item FGNWS can have yields in the 1 to 100~\emph{tons} gap which today separates conventional form nuclear weapons.
\item Compared to previous generations, FGNWs have enhanced direct coupling to dense targets and reduced collateral effects, as well as the capability to drive powerful ``jets'' and ``forged fragments.''
\item FGNWs are in line with the ``increased precision'' and ``reduced collateral damage'' trends of modern warfare.
\item Proponents will claim that FGNWs have a high-potential to destroy ``biological and chemical'' weapons, as well as future ``nanorobots...''
\item The military ``sweetness'' of FGNWs is such that they will eventually be built by all technologically-advanced countries, including non-nuclear weapons States such as Japan, Germany, Brazil, etc.
\end{enumerate}

\subsection{Technical aspects}
\label{Sec.7.2}

\begin{enumerate}
\item The development of FGNWs-related technologies does not require any ``political decision'' or ``conspiracy:'' The current \emph{dynamics of technology} is enough. (E.g., scientific competition, priority to ``hard'' technologies, aggressive commercialism, priority to military or ``dual'' technologies, etc.)
\item Even if FGNWs could be ``very far in the future,'' the disarmament dilemma is that it is precisely this ``remoteness'' which is taken as the excuse for developing the technologies that will ultimately lead to their construction.
\item Apart from inertial confinement fusion (ICF), the main areas of science and technology leading to the development of FGNWs are microelectromechanical engineering and nanotechnology.
\item The most promising technology for providing the high energy-density necessary to trigger very compact FGNWs is antimatter. 

\item As the main source of yield in FGNWs is the $DT$ fusion reaction, large-scale (i.e., kg level) tritium technology is required to deploy a full scale arsenal of FGNWs.  This implies that the deployment of FGNWs is linked to that of thermonuclear fusion reactors (See Ref.~\cite{GSPON2004A}.)
\item The main alternative source of tritium is its production by means of high-energy particle accelerators, which is at the same time the most matured technology for the large scale production of antimatter \cite{GSPON1987-, GSPON1983-, GSPON1997B}.
\end{enumerate}

\subsection{Political aspects}
\label{Sec.7.3}

\begin{enumerate}
\item As the United States are likely to take the lead in the development and deployment of FGNWs, a nuclear arms race may result in which all advanced industrialized countries (e.g., Germany, Japan, France, China, etc.) could compete to become the second largest military power in the XXIth century.
\item The development of FGNWs will most likely increase the attractiveness of previous generation nuclear weapons to technologically less-advanced countries. 
\item The most controversial aspect of $DT$-fusion-based FGNWs could be their use as anti-personnel weapons against unprotected soldiers or people.
\item Finally, proponents will argue that FGNWs's technologies may have significant medical, practical, and scientific applications.  While this could be true, one can also argue that such ``spin offs'' of primarily military technologies are not sufficient to justify a new major step in the arms race.
\end{enumerate}

\section*{Acknowledgements}
\label{ackno}

This paper is an opportunity to acknowledge the work done in the 23 years of existence of ISRI by many people who devoted a lot of time and effort to the scientific analysis of new types of nuclear weapons in order to open a well informed debate on their prospect.

\end{document}